\documentclass[sigconf,screen,nonacm]{acmart} 

\usepackage{amsfonts,balance,bbold,bm,color,enumitem,graphicx,hyperref,mathtools,microtype,multirow,natbib,soul,subcaption,tabularx,xcolor,xspace}
\usepackage[linesnumbered,ruled,vlined]{algorithm2e}
\usepackage{pgfplots}
\pgfplotsset{compat=1.18} 

\newcommand{\sn}{\textbf{\textit{SweetFruit}}\xspace}
\newcommand{\snn}{{\textit{SweetFruit}}\xspace}

\newcommand{\revisedtext}[1]{{\color{black} #1}}

\AtBeginDocument{%
  \providecommand\BibTeX{{%
    \normalfont B\kern-0.5em{\scshape i\kern-0.25em b}\kern-0.8em\TeX}}}

\def\USEACMSTYLE{}

\renewcommand\footnotetextcopyrightpermission[1]{}

\begin{document}

\title[\sn]{\sn: A Two-Stage Mobile Sensing System for Real-Time Fruit Sugar Estimation}

\date{March 2026}

\author{Mark Cardamis}
\email{m.cardamis@unsw.edu.au}
\orcid{0000-0001-7896-5038}
\affiliation{%
  \institution{University of New South Wales}
  \department{School of Computer Science and Engineering}
  \city{Sydney}
  \state{NSW}
  \country{Australia}
}

\author{Yanxiang Wang}
\email{yanxiang.wang@unsw.edu.au}
\orcid{0000-0002-1466-4006}
\affiliation{%
  \institution{University of New South Wales}
  \city{Sydney}
  \state{NSW}
  \country{Australia}
}

\author{Chun Tung Chou}
\email{c.t.chou@unsw.edu.au}
\orcid{0000-0003-4512-7155}
\affiliation{%
  \institution{University of New South Wales}
  \city{Sydney}
  \state{NSW}
  \country{Australia}
}

\author{Wen Hu}
\email{wen.hu@unsw.edu.au}
\orcid{0000-0002-4076-1811}
\affiliation{%
  \institution{University of New South Wales}
  \city{Sydney}
  \state{NSW}
  \country{Australia}
}

\renewcommand{\shortauthors}{Cardamis et al.}

\begin{abstract}
Accurate prediction of fruit sugar content is essential for quality control and market valuation in agriculture. Conventional measurement techniques rely on destructive, time-consuming processes (e.g., juicing and refractometry) or direct contact instruments, which hinder high-throughput operations. This paper introduces \textit{SweetFruit}, a mobile two-stage system that leverages low-cost sensors to estimate fruit sugar content without contact. In Stage~1, we implement a lightweight 3D deep learning model (\textit{SF-PointNet}) that uses point clouds from a Time-of-Flight (ToF) depth camera to classify fruit as high or low sugar. In Stage~2, a regression network (\textit{SF-Net}) predicts the fruit’s Brix value using measurements from a compact 18-channel near-infrared (NIR) spectrometer. The system uses simple off-the-shelf sensors (AS7265x NIR and Arducam ToF) with efficient processing pipelines for real-time execution on embedded platforms. Experiments on green `Granny Smith' apples and strawberries demonstrate the system’s effectiveness. Stage~1 achieves over 90\% classification accuracy, enabling rapid prescreening, while Stage~2 delivers precise sugar estimates, with a root mean square error (RMSE) of 0.57~\textdegree Brix, reducing error by 22\% compared to using NIR sensing alone. SweetFruit offers a scalable, field-ready solution for rapid fruit quality screening, showcasing the benefits of task-specific multimodal sensing in mobile agricultural applications.
\end{abstract}



\maketitle

\section{Introduction}

\ifdefined\USEIEEESTYLE
  \IEEEPARstart{T}{he} 
\else
  The 
\fi
sugar content in fruits is a key indicator of quality, influencing taste, consumer preference, and market value~\cite{fan_non-destructive_2020}. Sugar levels are typically measured in terms of Soluble Solids Content (SSC) or Brix value, which is the percentage of dissolved sugar content in an aqueous solution~\cite{grabska_analyzing_2023}. However, current sugar measurement techniques are limited by their invasive nature, high costs, or complex operational requirements, presenting a significant barrier to scalable, on-the-spot quality assessment. Developing a non-contact, low-cost, and accurate solution for estimating sugar content can substantially enhance scalability and efficiency, offering a promising path forward for high-throughput fruit quality monitoring in agricultural and retail environments.

\revisedtext{Conventional techniques for measuring sugar content typically depend on destructive or contact-based approaches~\cite{medic_remotely_2024}. Destructive methods, such as using extracted juice with a refractometer or high-performance liquid chromatography (HPLC), provide highly accurate results, however as they are time-intensive only a small number of fruits can be sampled per batch. Contact-based optical devices, including the Atago PAL-HIKARi Pocket Brix Meter~\cite{atago_pocket_2024}, eliminate the need for juice extraction but still require physical contact with the fruit. This direct-contact limits scalability and makes such methods unsuitable for high-throughput operations. These drawbacks have motivated the development of more practical solutions that preserve accuracy while remaining non-contact and cost-effective.}
 
Near-infrared (NIR) spectroscopy has emerged as a promising technology for non-invasive, non-destructive measurement of sugar content in fruits~\cite{zeng_research_2024, zhao_flight_2023, grabska_analyzing_2023}. Unlike visible light, NIR wavelengths penetrate the skin of the fruit (2-4 mm~\cite{nicolai_nondestructive_2007}), interacting with internal compounds such as sugars, water and organic acids~\cite{yu_development_2016}. Sugar molecules exhibit characteristic absorption peaks in the NIR region, specifically at wavelengths around 880 nm, 940 nm, 980 nm, and 1064 nm \cite{hua_design_2017}. By analyzing the reflected light spectrum, it is possible to correlate specific spectral signatures with sugar concentration. Previous studies often relied on high-cost, laboratory-grade instruments~\cite{huang_stacking_2024}, or direct-contact with the fruit~\cite{zhao_flight_2023,tran_portable_2020,noguera_new_2022}, limiting their scalability and practicality. 

Prior research and theoretical analysis suggest that factors such as environmental variability and fruit surface properties may influence spectral measurements~\cite{vaudelle_influence_2015,askoura_multispectral_2016,lohner_determining_2021-1}. Although our experiments did not explicitly isolate these effects, we note the following considerations:

\textit{Spectral noise}. Ambient lighting fluctuations and other environmental conditions can introduce noise into spectral measurements, potentially obscuring subtle features related to sugar content. In high-dimensional spectral data, it is possible that predictive models might inadvertently incorporate some of these extraneous variations. To address this possibility, our design employs a robust framework, advanced noise reduction techniques, and effective preprocessing (e.g., SNV normalization), described further in Section~\ref{subsec:datapipelinestep1} below.

\textit{Surface variations}. Variability in the fruit’s skin properties, including differences in thickness, texture, and surface irregularities, can potentially affect the reflected NIR spectrum. To mitigate these potential effects, we integrate a depth camera to capture the fruit’s surface profile and derive a corresponding depth vector to complement the spectral data, described in Section~\ref{subsec:datapipelinestep2} below.

\revisedtext{
In this paper, we present \sn, a low-cost sensing system that estimates fruit sugar content in real-time using a two-stage architecture and commercial off-the-shelf (COTS) sensors. \snn uses a simple Time-of-Flight (ToF) depth camera and a compact near-infrared (NIR) spectrometer to assess surface geometry and internal spectral absorption, adopting a coarse-to-fine design methodology to optimize throughput in practical deployments. \textit{Stage~1} is a lightweight screening stage whose goal is to rapidly classify fruits as high-sugar or low-sugar and route select samples to \textit{Stage~2} model for fine-grained Brix prediction.
This coarse-to-fine strategy mirrors successful, resource-constrained designs in other sensor systems where a first-stage reduces the load on expensive second-stage processing~\cite{lattanzi_lightweight_2023,viola_rapid_2001}. In practice, the Stage~1 classifier is tuned for high recall so that Stage~2 is invoked primarily for fruits likely to benefit from high-precision analysis, thus improving overall throughput by avoiding likely low-sugar fruits. For instance, if only 20\% of fruits exceed the sweetness threshold, Stage~1 filtering can save ~80\% of the Stage~2 detailed scans, significantly improving throughput.}

In summary, this paper makes the following contributions:
\begin{itemize}
  \item We design and implement \snn, an integrated hardware-software system for non-destructive fruit sweetness measurement. The prototype combines a multi-channel NIR spectral sensor with a ToF depth camera, along with custom data processing pipelines, to enable real-time predictions of fruit sugar content.
  \item We develop a two-stage sensing pipeline comprising: (i) a Stage~1 ToF-based classification model (SF-PointNet) to identify high vs. low sugar fruits at distances up to 50~cm, and (ii) a Stage~2 multimodal regression model (SF-Net) that fuses NIR spectral data with depth-derived features for accurate Brix estimation at close range (up to 15~cm). This multi-modal approach leverages complementary sensor modalities to improve prediction performance and robustness.
  \item We conduct extensive experiments on two types of fruits (apples and strawberries) to evaluate the SweetFruit system. The results show that the Stage~1 classifier can effectively distinguish high-sugar fruits with around 90\% accuracy on both datasets (using leave-one-out cross-validation), validating Stage~1 as a reliable filter for fast screening. For fruits requiring precise analysis, Stage~2 delivers accurate sugar content predictions: on apples, the combined sensing approach achieves an Root Mean Square Error (RMSE) of 0.57~\textdegree Brix, which is about 22\% lower error than state-of-the-art (SOTA) methods. These findings demonstrate the practicality of our two-stage approach for improving throughput without sacrificing accuracy.
\end{itemize}

The rest of the paper is organized as follows. Section~\ref{sec:background} introduces background and preliminaries of NIR spectroscopy and Sugar Composition. Section~\ref{sec:methodology} details the design and implementation of the \snn system, including sensor fusion and model architecture. Section~\ref{sec:evaluation} presents the experimental evaluations conducted using real-world data. Section~\ref{sec:relatedwork} reviews the related work, and finally, Section~\ref{sec:conclusion} concludes the paper with a discussion of results and future directions.

\section{Background and Preliminaries}\label{sec:background}
\ifdefined\USEIEEESTYLE \noindent \fi
\subsection{NIR Spectroscopy for Sugar Content Estimation}\label{subsec:sugarbackground}

NIR spectroscopy is a widely used non-destructive technique for estimating the sugar content of fruits, leveraging its ability to detect molecular vibrations associated with specific chemical bonds. In fruits, sugar content predominantly comprises fructose, glucose, and sucrose, with the composition and concentration varying by species, variety, and ripeness. For instance, in apples, fructose generally constitutes the largest proportion, followed by sucrose and glucose.

NIR spectroscopy measures light absorption in the near-infrared region (700–2500 nm), where molecular overtone and combination vibrations occur. Sugar molecules in apples exhibit characteristic absorption bands in NIR wavelengths, primarily linked to C-H and O-H bonds~\cite{wang_grading_2022, fan_non-destructive_2020, yu_development_2016}. Notable absorption peaks are commonly observed around 880 nm and 940 nm~\cite{hua_design_2017, fan_non-destructive_2020}, which correspond to these vibrational modes. This capability makes NIR spectroscopy an effective tool for quantifying soluble solids content.

\begin{figure}[htbp]
  \centering
  \subfloat[NIR spectral response of apple juice with varying sugar concentrations\label{fig:backgroundspectabrixsolution}]{\includegraphics[height=0.38\linewidth]{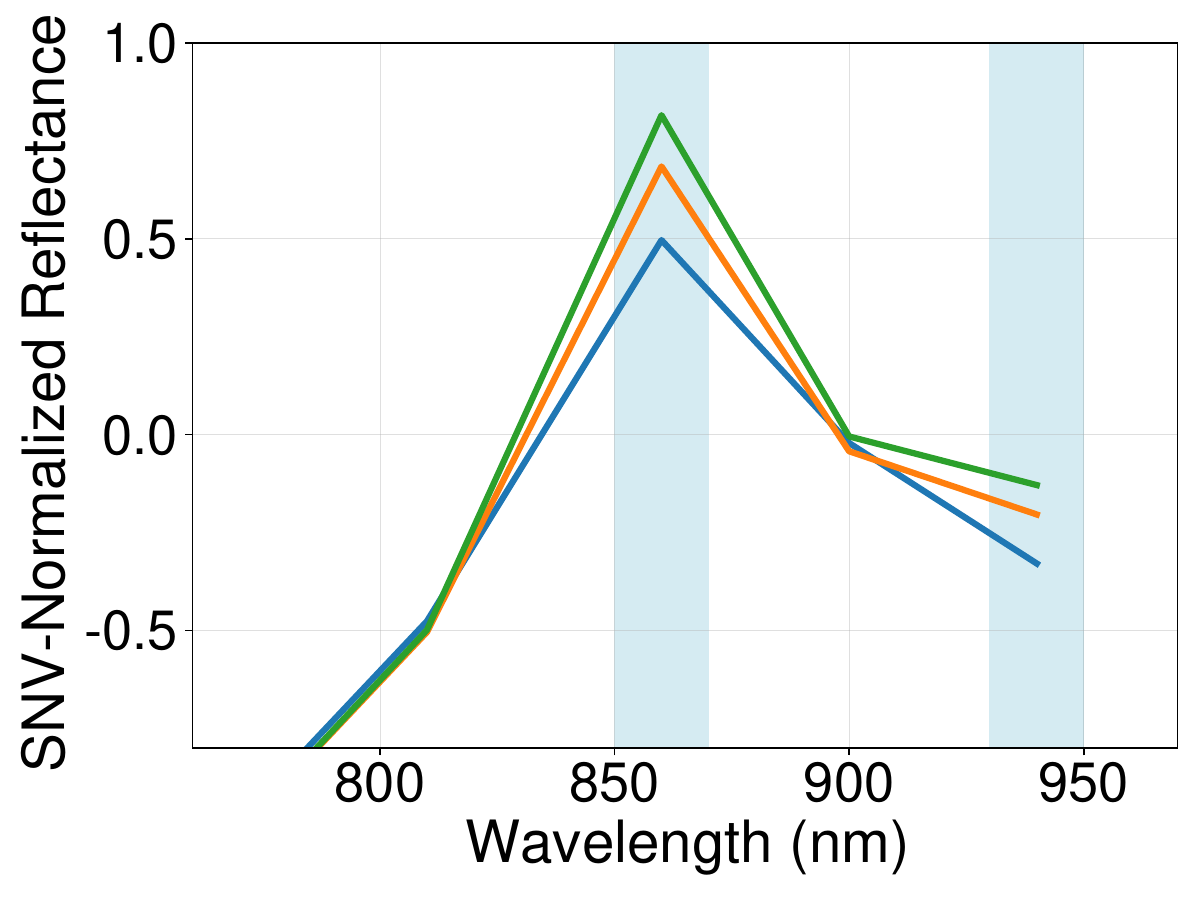}}
  \hspace*{0.5em}
  \subfloat[Light scattering for two-layer fruit model\label{fig:backgroundscatteringmechanisms}] {\includegraphics[trim={0cm -0.5cm 0.2cm 0.1cm},clip,height=0.35\linewidth]{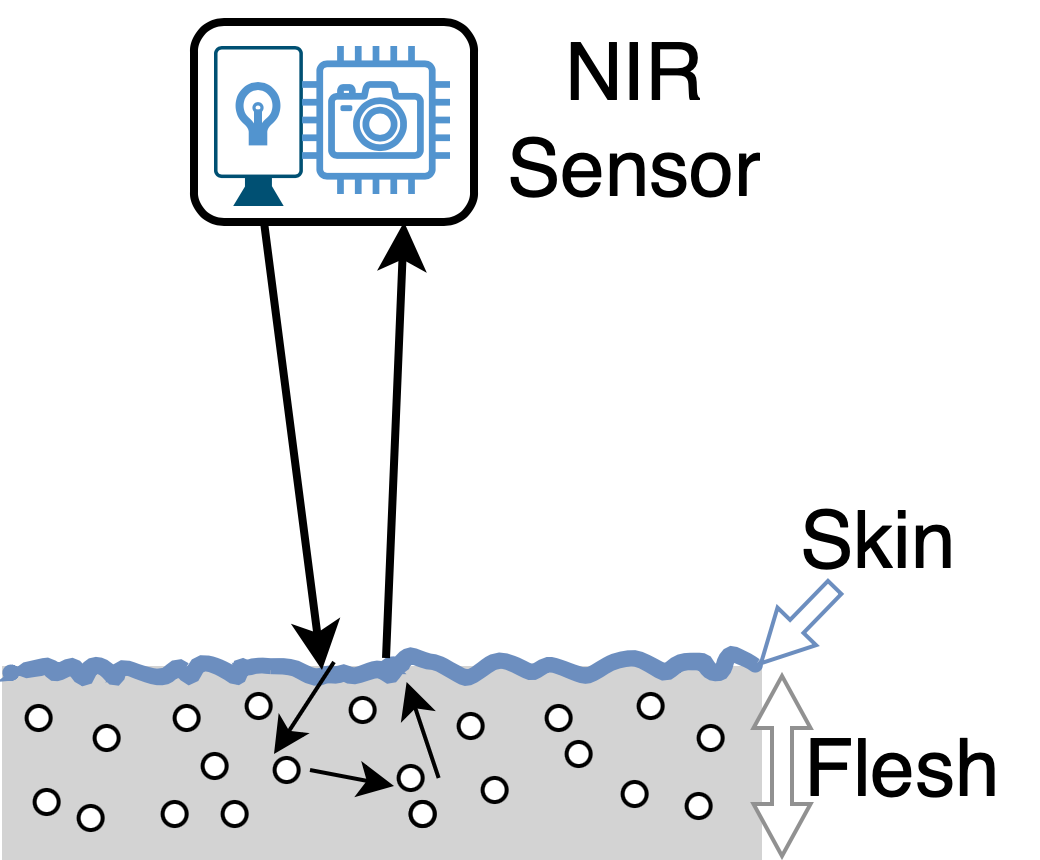}}
  \caption{Simplified NIR scattering concept for Brix.}
  \label{fig:backgroundnirspectroscopyofapples}
  \ifdefined\USEACMSTYLE
    \Description{}
  \fi
\end{figure}

A preliminary experiment was setup using 12\% Brix apple juice in a beaker, which was then water-diluted to 8\% and 4\%. Figure~\ref{fig:backgroundspectabrixsolution} illustrates the spectral response~\footnote{Multi-spectral measurements were captured using a Sparkfun AS7265x sensor~\cite{ams_as7265x_2024}} for different sugar concentrations, with absorption peaks observed in the NIR region at approximately 880 nm and 940 nm, consistent with prior studies~\cite{hua_design_2017, fan_non-destructive_2020}.  Furthermore, its ability to discern distinct spectral patterns for varying Brix levels demonstrates the reliability of NIR spectroscopy for accurately quantifying sugar concentrations.

\subsection{Fruit Composition and NIR Scattering Mechanisms}\label{subsec:applescattering}

NIR light interacts with fruits through complex optical processes influenced by the fruit's structural and compositional heterogeneity~\cite{fan_non-destructive_2020, tanaka_material_2019}. When NIR light is applied to a fruit, two main scattering mechanisms occur~\cite{askoura_multispectral_2016}: (i) surface scattering, where light reflects off the skin without penetrating the flesh, and (ii) subsurface scattering, where light enters the tissue, interacts with its structure, and then scatters back, as shown in Figure~\ref{fig:backgroundscatteringmechanisms}. The behavior of these scattering mechanisms is primarily governed by the scattering coefficient ($\mu_s$), which reflects structural properties like cell arrangement and density. Additionally, the absorption coefficient ($\mu_a$) corresponds to the concentrations of key chemical components, such as sugars, water, and pigments. Together, these coefficients provide insight into both the physical structure and chemical composition of the tissue~\cite{han_comb-based_2022}. 

In our study, we employ two complementary optical techniques to extract detailed information about the fruit's material properties. The NIR sensor captures spectral reflectance, directly revealing how variations in chemical makeup and tissue structure influence the intensity and distribution of reflected light. In parallel, the ToF camera provides spatially-resolved depth measurements that capture minute distance variations across the fruit’s surface. These depth variations encode information about light propagation through the tissue, linking structural characteristics to optical responses. By integrating reflectance data from the NIR sensor with depth information from the ToF camera, our approach enhances material classification by leveraging both chemical and structural attributes of the fruit tissue~\cite{han_comb-based_2022, nicolai_nondestructive_2007}.

\subsection{ToF Cameras and Distance Calculation}\label{subsec:tofdistancecalculation}

ToF cameras illuminate a scene with a modulated light source, such as a laser or LED, operating at a specific frequency. The reflected light from objects in the scene is captured by the camera's sensor. By comparing the phase of the emitted light with that of the received light, the camera calculates the phase shift, which is directly related to the distance traveled by the light~\cite{li_time--flight_2014}.

The phase shift (\(\phi\)) is determined using multiple measurements of the reflected light, each taken with a different reference phase. Typically, four measurements are taken with reference phases offset by 90 degrees. Let \(Q_1\), \(Q_2\), \(Q_3\), and \(Q_4\) represent the measured values at these phases. The phase shift is calculated as:

\[
\phi = \arctan{\left(\frac{Q_3 - Q_4}{Q_1 - Q_2}\right)}
\]

Once the phase shift is known, the distance (\(d\)) to the object can be calculated using the formula:

\[
d = \frac{c}{4\pi f} \phi
\]

where: $c$ is the speed of light and $f$ is the modulation frequency of the emitted light.

This equation is used to translate the raw camera frame into a depth image which is used later in Section~\ref{subsec:datapipelinestep2}.

\section{Methodology}\label{sec:methodology}
\ifdefined\USEIEEESTYLE \noindent \fi

\begin{figure*}[htbp]
  \centering
  \includegraphics[trim={0.1cm 0.2cm 0.1cm 0cm},clip,width=0.8\linewidth]{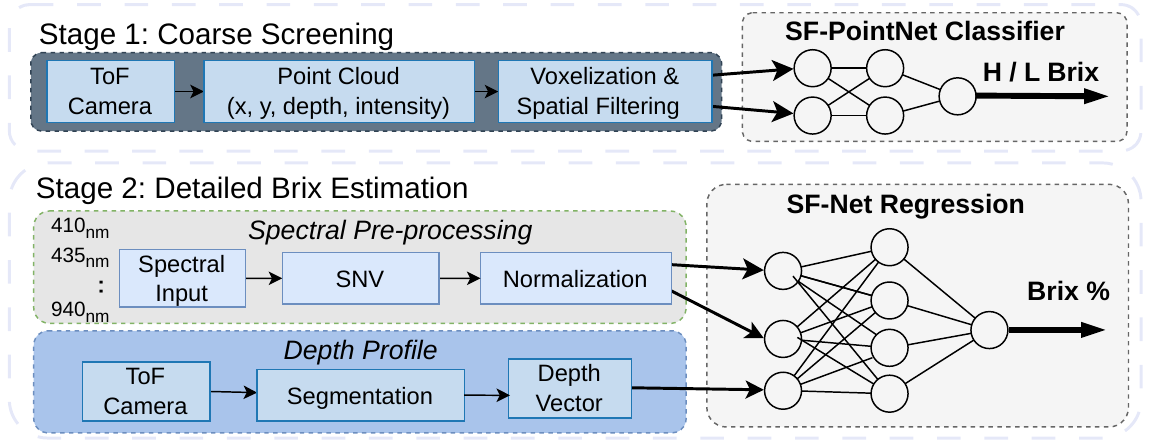}
  \caption{Overview of the \snn workflow. \revisedtext{Stage~1 uses a voxelized point cloud fed into a lightweight PointNet-inspired network to output a binary high/low sugar classification. Stage~2 takes the 18-dimensional spectral feature (after SNV normalization) concatenated with a 10-value depth feature vector, passing them through fully connected layers to output a Brix prediction}.}
  \label{fig:methodoverviewtwostagesystemworkflow}
  \ifdefined\USEACMSTYLE
    \Description{Two-stage sugar estimation pipeline integrating SF-PointNet and SF-Net models.}
  \fi
\end{figure*}

To optimize the performance and throughput of \snn, we design a two-stage sugar content estimation framework, the overall workflow is shown in Figure~\ref{fig:methodoverviewtwostagesystemworkflow}. In \nameref*{subsec:datapipelinestage1}; a coarse 3D scan of the fruit is taken at far distances (>16 cm) using a ToF depth camera. In \nameref*{subsec:datapipelinestage2}; a closer-range scan (<16 cm) obtains the spectral and depth features fused by SF-Net model for precise Brix estimation. We describe each stage below.

\subsection{Stage 1: ToF-Based Rapid Classification}\label{subsec:datapipelinestage1}

\begin{figure}[htbp]
  \centering
  \includegraphics[trim={0.7cm 0.1cm 0.7cm 0.1cm},clip,width=1\linewidth]{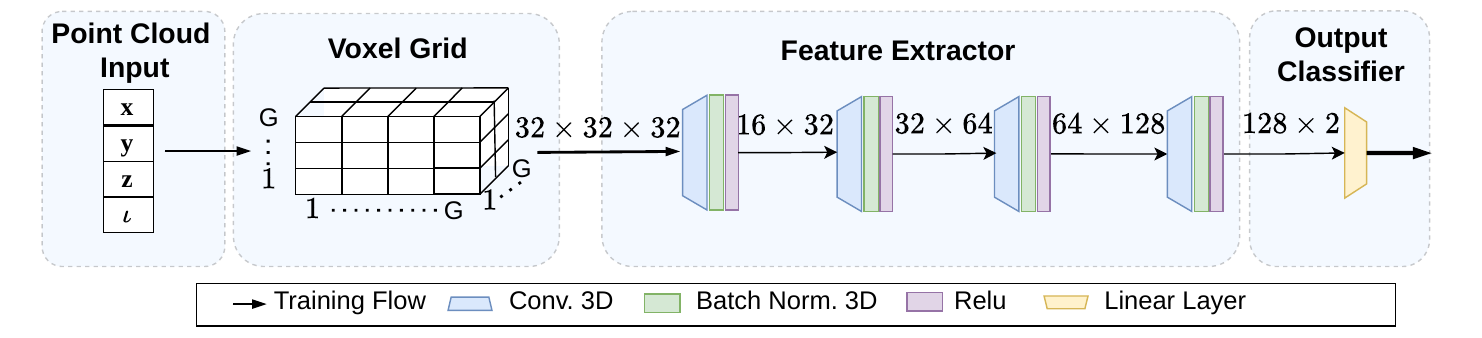}
  \caption{Stage~1 point cloud classification network (SF-PointNet). The 3D fruit point cloud is voxelized into a $32^3$ grid with multiple feature channels (intensity, etc.). Four 3D Conv. layers (Conv3D + BN + ReLU) capture shape features. Global pooling and a fully-connected layer produce a binary high/low sugar prediction.}
  \label{fig:methodpointnextarchitecture}
  \ifdefined\USEACMSTYLE
    \Description{A block diagram representing the architecture of the SF-PointNet deep learning model.}
  \fi
\end{figure}

Stage~1 performs a rapid, non-contact screening of fruit sweetness using depth-sensor data alone. Low-resolution three-dimensional spatial measurements $(x, y, z, \text{intensity})$ acquired from an Arducam ToF camera are processed by SF-PointNet, a lightweight 3D convolutional neural network, to classify fruits into high/low sugar categories based solely on ToF-derived point cloud features (Figure~\ref{fig:methodpointnextarchitecture}). To standardize input complexity and improve robustness, raw point clouds are voxelized and preprocessed through region segmentation and spatial filtering, as described in Section~\ref{subsubsec:highintensityregionextraction}. Prediction confidence thresholds and rejection sampling are applied to manage classification uncertainty, enabling reliable selective screening. The resulting binary outputs are subsequently used to prioritize high-sugar fruits for Stage~2, thereby reducing unnecessary close-range scanning.

\subsubsection{Fruit Region Segmentation with Spatial Radius Constraints}\label{subsubsec:highintensityregionextraction}

To accurately segment individual fruits from the acquired ToF depth images, we adopt a region-growing segmentation approach that prioritizes spatial uniformity and controlled depth variation. This technique effectively segments fruit regions while mitigating the influence of minor surface irregularities and background noise.

Given an intensity (confidence) map \( I \) and its corresponding depth map \( D \), the algorithm identifies an initial seed point at the pixel with the highest intensity. Beginning from this seed, the segmentation process iteratively expands outwards by evaluating neighboring pixels based on two primary constraints. First, a spatial constraint is enforced such that neighboring pixels are only included if their Euclidean distance from the seed point \((x_0,y_0)\) remains within a specified maximum spatial radius \( R_{\text{max}} \). This constraint promotes a circular segmentation pattern consistent with typical fruit geometry. Second, the algorithm incorporates a depth constraint, restricting significant depth variations within the segmented region. Specifically, a candidate pixel at location \((x',y')\) neighboring a pixel at \((x,y)\) is included only if its intensity value satisfies:

\begin{equation}
    I(x',y') < I(x,y) \left[ 1 + w_{D}\frac{|D(x',y') - D(x,y)|}{\Delta D_{\text{max}}} \right],
\end{equation}
subject to the spatial radius constraint:
\begin{equation}
    \sqrt{(x'-x_0)^2 + (y'-y_0)^2} \leq R_{\text{max}},
\end{equation}
where \((x_0,y_0)\) is the seed point, \( R_{\text{max}} \) represents the maximum allowed spatial radius (e.g., 30 pixels), \( w_{D} \) is a depth weighting factor (0.05), and \(\Delta D_{\text{max}}\) specifies the maximum permissible depth difference (100\,mm). 

This iterative region-growing continues until no additional neighboring pixels fulfill these spatial and depth conditions. The resulting segmentation is subsequently refined using morphological operations to fill minor holes, producing a coherent and continuous binary mask that accurately outlines each fruit. This segmentation method provides stable, geometrically consistent regions suitable for subsequent classification and detailed spectral analyses.

\subsubsection{SF-PointNet Deep Learning Model}

\paragraph{Point Cloud Preprocessing} From the segmented fruit region, we obtain a set of 3D points ${(x,y,z)}$ for the fruit surface. We transform these points into a local fruit-centered coordinate frame: the point cloud is translated such that the fruit’s centroid is at the origin, and we scale the coordinates to a fixed bounding radius. We then rasterize the points into a 3D grid of size $G \times G \times G$ (we use $G=32$ in our experiments). Each voxel accumulates a set of points (possibly zero); we compute per-voxel features which serve as input channels to the network. To handle varying fruit sizes and point counts, we pad or truncate the point cloud to a fixed number of points before voxelization (ensuring consistent tensor shape). The result is a $G^3 \times C$ tensor (with $C=4$ channels in our default configuration) representing the fruit’s 3D profile.

\paragraph{Network Architecture} Our Stage~1 classifier is a 3D CNN network derived from modern point-based models, figure~\ref{fig:methodpointnextarchitecture} illustrates the architecture. The voxel grid is first passed through a series of four 3D convolutional blocks. Each block uses a 3×3×3 convolution followed by BatchNorm and ReLU activation. We increase the number of channels in deeper blocks to learn higher-level features; using blocks output 16, 32, 64, and 128 feature maps respectively. The resulting 128-channel feature volume is global average pooled to produce a 128-D feature vector for the entire fruit. Finally, a fully-connected layer (with softmax activation) produces two outputs: the probability of the fruit being low-sugar (class~0) or high-sugar (class~1).

\paragraph{Training Details} Given the limited dataset size, we train the Stage~1 classifier using leave-one-out cross-validation. The loss function is binary cross-entropy. However, to account for any class imbalance, we employ a focal loss variant with $\gamma=2.0$ and $\alpha=0.75$ (focusing the learning on harder, misclassified examples). We train using the Adam optimizer (initial learning rate $6\times10^{-4}$) with a cosine decay schedule (minimum LR 5\% of initial) and a linear warm-up for the first 3 epochs. Each training fold runs for up to 60 epochs with early stopping (patience 8, monitoring validation F1 score for the positive class). We reserve 15\% of the training set in each fold as a validation split to determine the best epoch and to tune the classification threshold if needed. At inference, we use the default 0.5 probability threshold for binary decisions.

\begin{figure}[htbp]
  \centering
  \subfloat[Robot arm with sensors\label{fig:methodrobotarmfront}]{\includegraphics[width=0.46\linewidth]{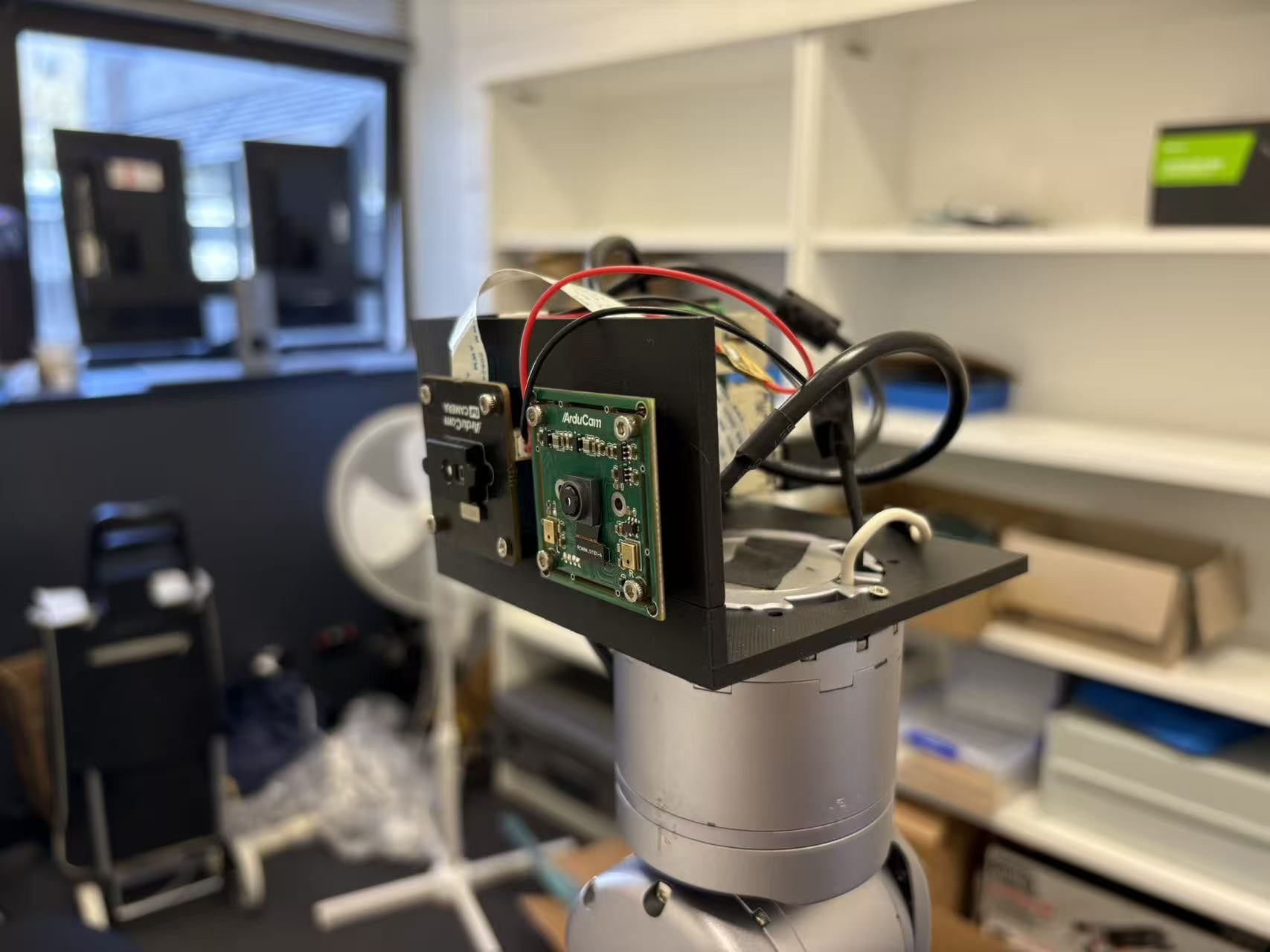}}
  \hspace*{0.5em}
  \subfloat[Multiple fruits classified with single Stage~1 scan \label{fig:methodmultiscanfruitclassification}]{\includegraphics[width=0.5\linewidth]{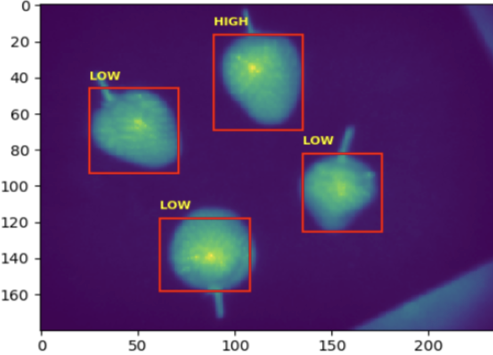}}
  \caption{Practical deployment for two-stage pipeline.}
  \label{fig:methodstage1robotarm}
  \ifdefined\USEACMSTYLE
    \Description{}
  \fi
\end{figure}

\revisedtext{
\subsubsection{Practicality and throughput analysis}
To justify the two-stage pipeline we quantify both (i) per-sample acquisition/processing time and (ii) expected throughput when Stage~1 is used as a selective filter.

\paragraph{Measurements and assumptions} On our prototype the ToF camera streams at up to 30\,fps while the NIR spectrometer returns a spectral reading in approximately 1\,s (depending on averaging and integration settings). In our initial experiments we collected 4,000 ToF frames per fruit to build robust point-cloud representations (see Section~\ref{subsubsec:highintensityregionextraction}); this was intended to create a high-quality evaluation dataset rather than to represent an optimized deployment capture strategy.

\paragraph{Optimized operational regime} For practical deployments we propose the following reduced acquisition strategy:
\begin{itemize}
  \item Stage~1 (Coarse Screening): acquire a single ToF frame of the fruit, which can be captured in <0.1\,s at 30\,fps, for Brix classification using SF-PointNet model.
  \item Stage~2 (Detailed Brix Scan): reposition the sensors, using a robot arm as shown in Figure~\ref{fig:methodrobotarmfront}, to the optimal NIR distance (<15 cm) $\approx$ 2\,s, then capture a single NIR spectral reading in <1\,s for precise Brix estimation using SF-Net model; total Stage~2 acquisition per fruit is therefore typically $\approx$ 3\,s.
\end{itemize}

\paragraph{Throughput example} Under the optimized strategy above and assuming Stage~1 simultaneously scans at least 4 fruits (e.g. Figure~\ref{fig:methodmultiscanfruitclassification}) and only accepts the top 20\% of fruits for Stage~2 scan, expected per-fruit average time is:
\[
T_{\text{avg}} = T_{\text{Stage1}} + p_{\text{accept}} \cdot T_{\text{Stage2}},
\]
where \(T_{\text{Stage1}}\!\approx\!0.1\text{ s}\), \(T_{\text{Stage2}}\!\approx\!3\text{ s}\), and \(p_{\text{accept}}\!=\!0.20\) gives \(T_{\text{avg}}\!\approx\!0.7\text{ s}\). These numbers illustrate the order-of-magnitude throughput advantage of selective scanning versus running Stage~2 on every sample.
}

\subsection{Stage 2: High-Resolution NIR–Depth Brix Regression}\label{subsec:datapipelinestage2}

Stage~2 provides a precise fruit sugar content estimation by fusing spectral data with depth data. We refer to this multimodal input as depth-encoded spectral measurements, which compensates for inconsistencies in spectral reflections caused by fruit surface irregularities, ensuring robust and accurate predictions.

We design a custom deep learning model, called \textit{SweetFruit-Net (SF-Net)}, to estimate the fruit’s Brix value from these multimodal inputs. First, the 18-dimensional NIR spectral vector is preprocessed (e.g., applying standard normal variate normalization to reduce effects of lighting and sensor drift). In addition, a feature vector describing the fruit’s spatial properties is extracted from the depth map. The spectral and depth feature vectors are then concatenated to form a combined input for SF-Net.

\subsubsection{Signal Pre-processing}\label{subsec:datapipelinestep1}

The first step in the \snn Stage 2 workflow involves pre-processing the raw spectral data to mitigate the impact of noise and variability introduced by environmental factors, fruit surface properties, and sensor limitations.

\paragraph{Scattering correction}
Scattering correction mitigates distortions in spectral data caused by baseline shifts and light scattering on uneven surfaces~\cite{zhao_flight_2023}. These distortions can arise from variations in surface roughness, fruit texture, and environmental conditions, introducing variability into spectral measurements. Standard Normal Variate (SNV) normalization is applied to each spectrum to standardize intensity values, as follows: 

\[
x_i^{\text{SNV}} = \frac{x_i - \bar{x}}{\sigma_x}
\]
where $x_i$ is the raw intensity at the $i\text{-th}$ wavelength, $\bar{x}$  is the mean intensity of the spectrum and $\sigma_x$ is the standard deviation. This helps ensure that variations in spectral response are primarily attributed to compositional differences rather than extraneous influences.

\begin{figure}[htbp]
  \centering
  \subfloat[No preprocessing\label{fig:methodspectranoprocessing}]{\includegraphics[width=0.48\linewidth]{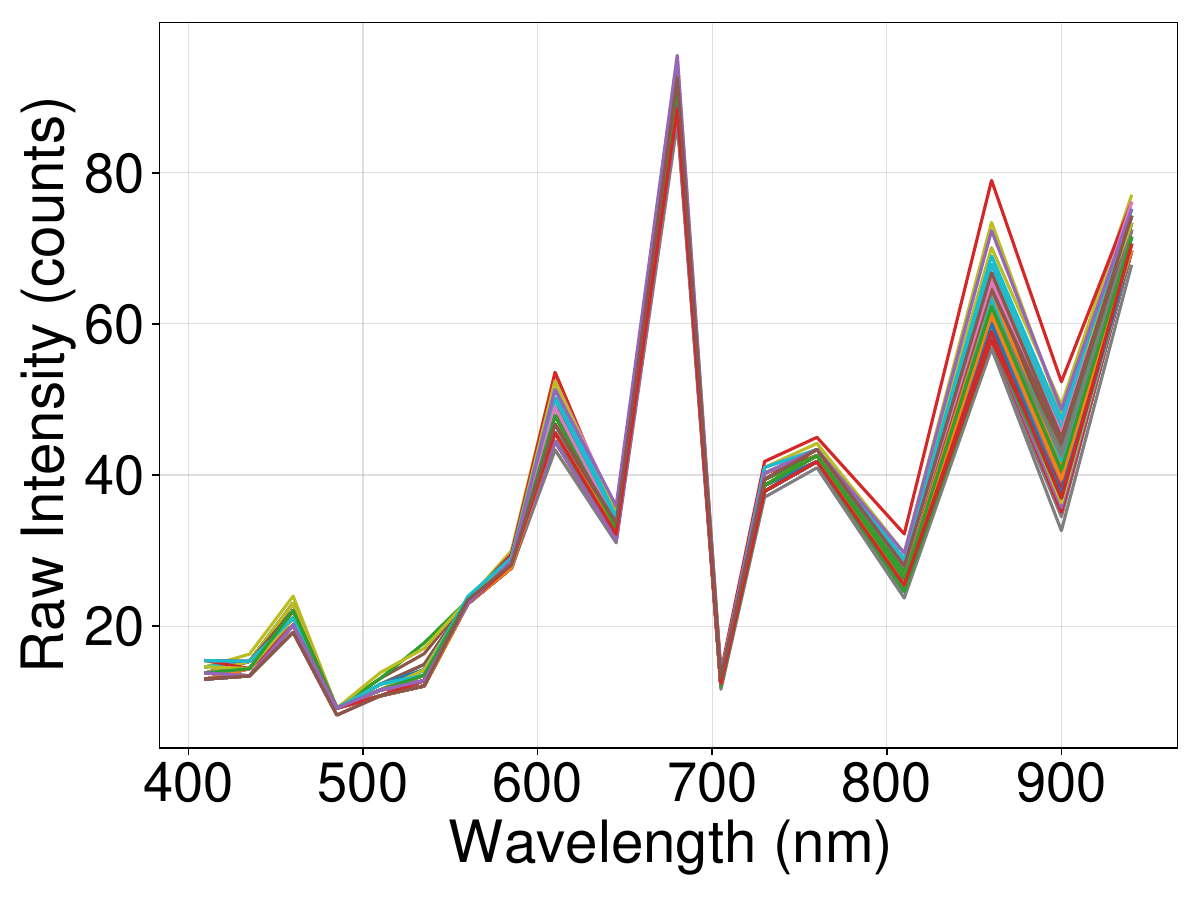}}
  \hspace*{0.5em}
  \subfloat[SNV preprocessing\label{fig:methodspectrasnvprocessing}]{\includegraphics[width=0.48\linewidth]{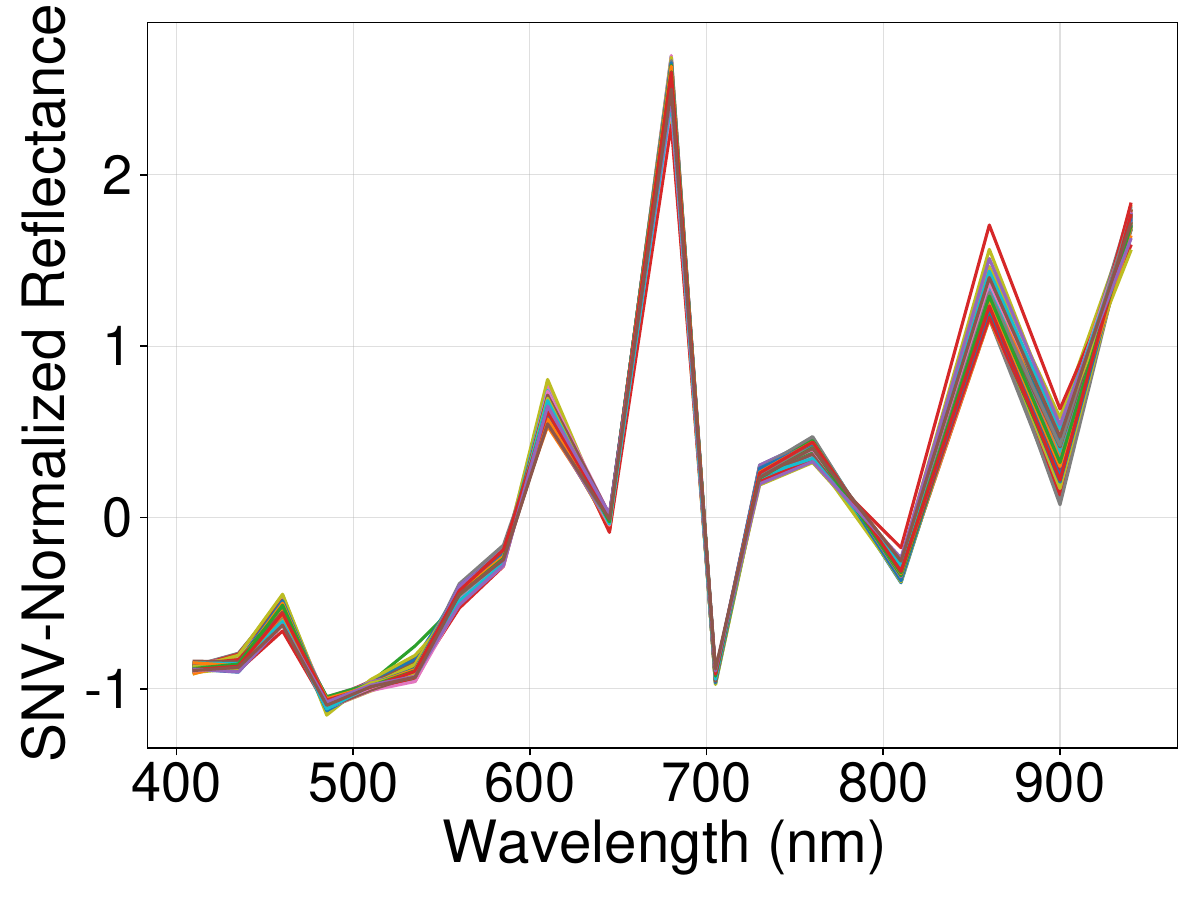}}
  \caption{Spectral responses effects of preprocessing.}
  \label{fig:methodspectralresponseswithpreprocessing}
  \ifdefined\USEACMSTYLE
    \Description{}
  \fi
\end{figure}

Figure~\ref{fig:methodspectralresponseswithpreprocessing} illustrates the effectiveness of SNV normalization in aligning absorption peaks and improving comparability across spectra by standardizing intensities.. The processed data serves as a reliable foundation for subsequent analysis and processing.

\paragraph{Normalization}

While SNV processing addresses scattering effects and baseline shifts within individual spectra, it does not fully account for broader inconsistencies caused by environmental variations such as sensor drift or fluctuating lighting across samples. To complement SNV, an additional normalization step ensures consistency across spectra by scaling intensity values to a range of 0 to 1: 

\[
x_i^{\text{norm}} = \frac{x_i - x_{\text{min}}}{x_{\text{max}} - x_{\text{min}}}
\]
where $x_{\text{min}}$ and $x_{\text{max}}$ are the minimum and maximum intensities in the spectrum. This step minimized inter-sample inconsistencies, providing a standardized input for subsequent processing.

\subsubsection{Depth Profile Segmentation}\label{subsec:datapipelinestep2}

The depth profile is captured using a Continuous Wave (CW) ToF camera, which emits modulated infrared light and measures the reflected signal to calculate object distances. This generates a depth map that highlights surface properties such as curvature and texture.

\begin{figure*}[htbp]
  \centering
  \subfloat[3D Point Cloud \label{fig:methodapplepointcloud}]{\includegraphics[trim={2cm 3.3cm 2cm 2cm},clip,width=0.38\linewidth]{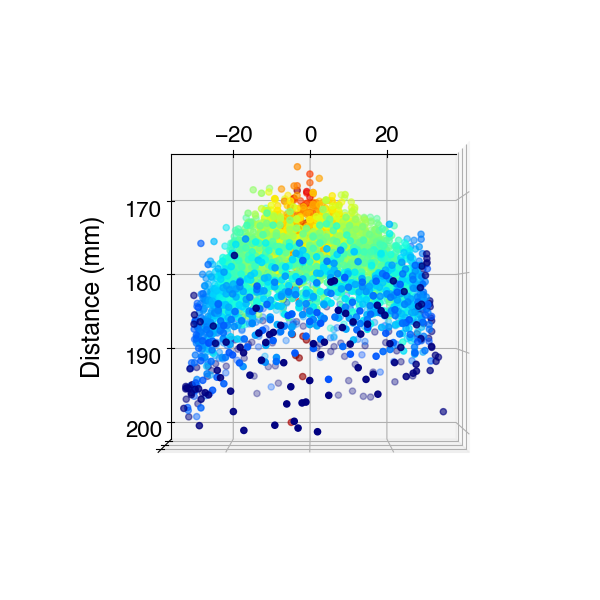}}
  \hspace*{0.3em}
  \subfloat[Rate of Depth vector change. \label{fig:methoddepthchangewindowsizes}]{\includegraphics[trim={0cm 0.8cm 0cm 0cm},clip,width=0.5\linewidth]{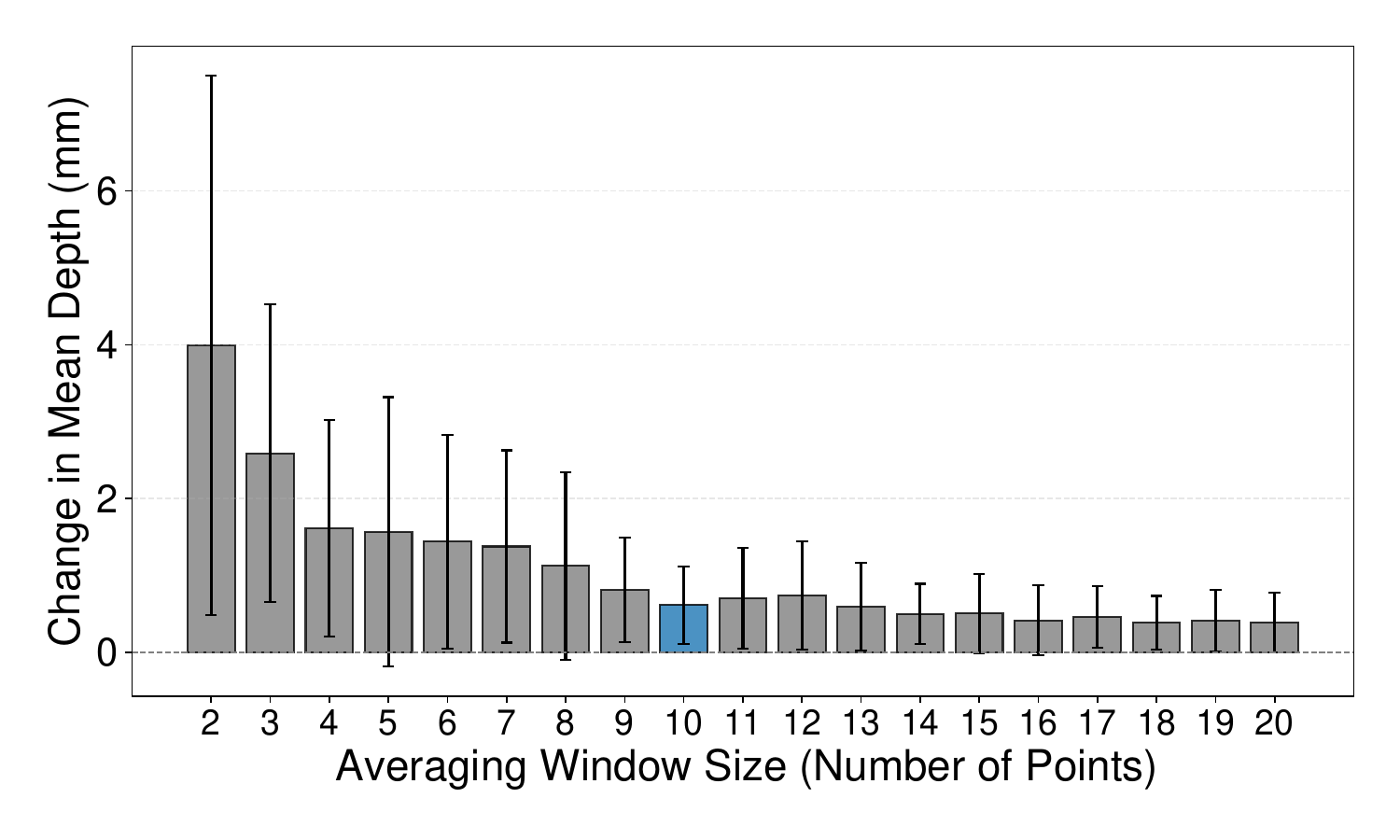}}
  \caption{Extracting the ToF depth profile.}
  \label{fig:methodappledepthprofile}
  \ifdefined\USEACMSTYLE
    \Description{}
  \fi
\end{figure*}

To isolate the fruit from the background, an intensity threshold is applied to the ToF camera image. The segmented depth image is then translated into a 3D point cloud (Figure~\ref{fig:methodapplepointcloud}), representing the fruit's surface geometry. This segmented profile serves as the basis for depth-encoded spectral adjustments.

\paragraph{Depth-Driven Spectral Adjustments}

To address potential effects arising from structural heterogeneity and surface irregularities in the fruits, the system incorporates depth-encoded spectral refinement. This process involves obtaining a depth profile of the fruit using a ToF camera and integrating this information with spectral data to enhance compositional analysis.

From the segmented depth image, the system extracts the 10 highest-intensity depth points to approximate local surface variability. This lightweight approach captures prominent surface features while reducing noise and maintaining computational efficiency. Figure~\ref{fig:methoddepthchangewindowsizes} illustrates how increasing the averaging window size smooths depth variations, stabilizing depth representation but potentially diminishing sensitivity to localized changes. 

By encoding these depth variations into a structured vector, the model enhances robustness against structural inconsistencies in the fruit’s surface. This depth-encoded spectral refinement mitigates the effects of skin texture and curvature on spectral readings, ultimately improving prediction accuracy in non-contact sugar content estimation.

The depth-encoded spectral refinement proceeds as follows:
\begin{enumerate}
    \item The ToF camera captures the depth profile of the fruit and generates a segmented depth image.
    \item From this segmented image, the 10 highest-intensity pixels are identified and used to construct the depth vector.
    \item The depth vector is concatenated with the preprocessed spectral data (including SNV and normalization adjustments).
    \item The combined input is fed into SF-Net model for Brix prediction.
\end{enumerate}

\subsubsection{SF-Net Deep Learning Model}\label{subsec:datapipelinestep3}

\begin{figure}[htbp]
  \centering
  \includegraphics[trim={0.6cm 0.2cm 0cm 0cm},clip,width=1.0\linewidth]{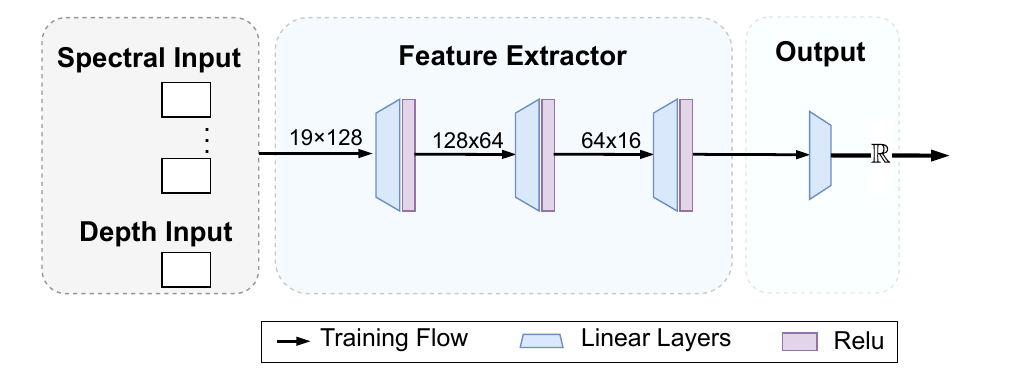}
  \caption{Architecture of SF-Net}
  \label{fig:methoddeeplearningmodel}
  \ifdefined\USEACMSTYLE
    \Description{A block diagram representing the architecture of the SF-Net deep learning model.}
  \fi
\end{figure}

The heart of the \snn system is the Sweet Fruit Network (SF-Net), a Fully Connected Neural Network designed to estimate sugar content from depth-encoded spectral data. Figure~\ref{fig:methoddeeplearningmodel} illustrates the architecture of SF-Net, which consists of several fully connected layers optimized for processing both spectral and depth features. The model is structured to handle the multi-channel complexity of spectral data while integrating depth information to improve prediction accuracy.

\paragraph{Model Architecture}
SF-Net comprises two main components: a feature extractor and a regression decoder. The feature extractor is implemented using a sequence of fully connected layers that progressively reduce the dimensionality of the input features. Each layer in the extractor applies ReLU activation to capture non-linear relationships in the spectral and depth data. This enables the network to identify patterns corresponding to sugar absorption bands in the spectral data, while simultaneously leveraging depth features to account for surface irregularities and structural variability. The model begins with an input layer that takes the concatenated spectral and depth data. It passes through three hidden layers with 128, 64, and 16 neurons, respectively, refining the extracted features for regression.
The regression decoder maps the extracted features to a single continuous output representing the estimated Brix value. This final layer ensures that the model produces a precise scalar prediction for each input sample. The regression process is optimized using a Mean Squared Error (MSE) loss function, which minimizes the squared difference between the predicted and actual Brix values:
    \[
    \text{MSE Loss} = \frac{1}{n} \sum_{i=1}^{n} (y_i - \hat{y}_i)^2,
    \]
    where \(y_i\) is the ground truth Brix value, and \(\hat{y}_i\) is the model's prediction.

\paragraph{Optimization and Regularization}
SF-Net incorporates several techniques to enhance generalization and mitigate overfitting. The weights of all layers are initialized using Xavier initialization, which ensures stable gradient flow during training. Additionally, L2 regularization, implemented as weight decay in the Adam optimizer, helps to prevent overfitting by penalizing large weight magnitudes. To adapt the learning rate dynamically, a ReduceLROnPlateau scheduler is employed. This scheduler monitors the validation loss and reduces the learning rate when no improvement is observed over a defined patience period, ensuring efficient convergence.

\paragraph{Training Workflow}
The training process follows a structured workflow. Each input batch, containing concatenated spectral and depth data, is passed through the fully connected layers of the model. At each layer, ReLU activations are applied to introduce non-linear transformations and enhance the model’s ability to capture complex patterns. The final output layer produces a scalar prediction for the Brix value, which is compared to the ground truth using the Mean Squared Error (MSE) loss function. The computed loss is then backpropagated through the network, and the optimizer updates the weights to minimize the loss. Validation metrics, including the validation loss, are recorded at the end of each epoch. These metrics guide the learning rate adjustments implemented by the scheduler, promoting stable and efficient training.

\section{Evaluation}\label{sec:evaluation}
\ifdefined\USEIEEESTYLE \noindent \fi
\subsection{Goals, Methodology and Metrics}\label{subsec:goalsMethodologyMetrics}

\subsubsection{Goals}
Our evaluation has three main objectives: first, assess the ability of the \snn system to predict the sugar content of fruits with high precision and reliability; second, benchmark \snn's performance against existing SOTA methods in terms of RMSE; and third, conduct an ablation study to evaluate the contributions of each system component, such as depth-aware spectral adjustments and preprocessing.

\subsubsection{Methodology}\label{subsubsec:EvaluationMethodology}

\paragraph{Prototype}
To evaluate \sn, we developed a low-cost prototype integrating the Sparkfun AS7265x~\cite{ams_as7265x_2024} multi-spectral sensor with the Arducam~\cite{arducam_time_2024} ToF depth camera. This system was optimized to target critical NIR wavelengths associated with sugar absorption, with the AS7265x sensor capturing 18 channels ranging from 410 nm to 940 nm and the Arducam ToF depth camera operating at 940 nm.

\paragraph{Experimental Setup}
The prototype was set up in an indoor environment to reduce external variability. The fruits were suspended from the ceiling using a thin cotton string and positioned in-line with the prototype, as shown in Figure~\ref{fig:methodappleindoorexperimentsetup}. 

For Stage~1 ToF-based measurements, apple and strawberry samples were placed at distances ranging from 16~cm to 40~cm from the sensor. Thresholds used for high/low sugar classifications were 13\% Brix for apples~\cite{wang_grading_2022} and 8.5\% Brix for strawberries~\cite{k_enhancing_2024,university_of_agronomic_sciences_and_veterinary_medicine_of_bucharest_romania_evaluation_2019}.

For Stage~2, apples were tested using night-time lighting at a distance of 15 cm $\pm$ 2 cm, with measurements taken simultaneously by the Sparkfun AS7265x sensor and the Arducam ToF camera. Each apple was measured once, with the side facing the sensors marked to be used in future ground-truth measurements. \revisedtext{For strawberries, NIR-only was used for Stage~2 regression with measurements taken by the AS7265x at a distance of 3 cm $\pm$ 1 cm. The AS7265x was not able to reliably measure the strawberry at the system's working distance (15\,cm) and the NIR operating distance of 3\,cm is outside the Arducam ToF range.}

\paragraph{Data Collection}
A total of 60 green apples and 30 strawberries were used for experimental testing, sourced from various local markets to ensure a realistic variability in sugar content and physical properties, as shown in Table~\ref{tab:sugar_content_apple_samples}. All samples were in a good condition with no bruising and similar ripeness. The Stage~1 model was trained using 4,000 ToF frames per fruit, each slowly rotated to generate robust geometric representations. Point clouds were voxelized and preprocessed using region segmentation and spatial filtering. Ground-truth labels are assigned using digital refractometer Brix values, described below.

\paragraph{Ground truth Measurements}
Following the non-invasive spectral and depth measurements, the sugar content of each fruit (apple and strawberry) was determined using a digital refractometer. For apples, a 1 cm deep segment was extracted from the equator, juiced, and filtered through a coffee filter to remove pulp and particulates. For strawberries, the leaves and stem were removed before juicing entire fruit. The filtered juice was then placed on the digital refractometer, and sugar content was recorded in \textdegree Brix.

\subsubsection{Metrics}
We used the RMSE of the Brix values to evaluate the performance of \sn. RMSE was calculated as
\[
\text{RMSE} = \sqrt{\frac{1}{n} \sum_{i=1}^{n} (y_i - \hat{y}_i)^2}
\]
where \(y_i\) represents the ground-truth sugar content (Brix), and \(\hat{y}_i\) represents the estimated sugar content. RMSE provides a clear measure of the system's predictive accuracy, with lower values indicating better performance. For training and testing our models, we used a leave-one-out cross-validation (LOOCV) approach \revisedtext{constructed at the fruit level. In each fold, we hold out every sample of one fruit (all spectral and ToF frames from that fruit) as the test set and train on the remainder.} This was repeated so that every sample was tested once and we report the average error across all LOOCV folds. \revisedtext{While our dataset is limited in size, LOOCV produces an unbiased per-fruit error estimate and reduces the chance of overfitting.}

\subsection{Stage 1 SF-PointNet Classification Performance}

For the both the apple and strawberry dataset, our Stage~1 classifier achieved strong performance in distinguishing high vs. low sugar fruits. Training the SF-PointNet classifier model using LOOCV (one fruit tested per iteration), the classifier attained an average accuracy of approx. 90\% for the high-sugar class, as shown below in Figure \ref{fig:resultspredictionbrixbinary}. In practical terms, this means 54 out of 60 apples were correctly classified into the proper sweetness category.

\begin{figure}[htbp]
  \centering
  \subfloat[Setup\label{fig:methodappleindoorexperimentsetup}]
  {\includegraphics[trim={2cm -1cm 2cm 2.5cm},clip,height=0.3\linewidth]{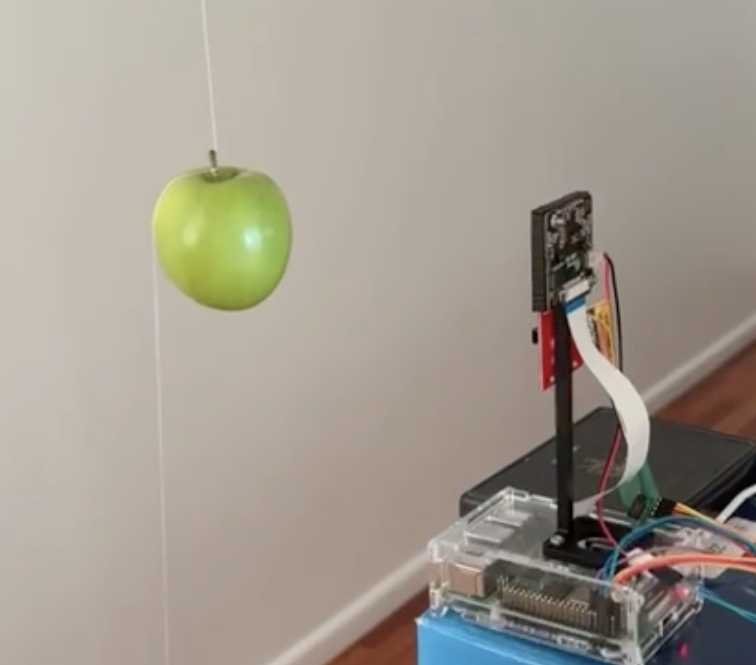}}
  \hspace*{0.2em}
  \subfloat[SF-PointNet\label{fig:resultspredictionbrixbinary}]{\includegraphics[trim={0cm 0.2cm 0.8cm 0.5cm},clip,height=0.3\linewidth]{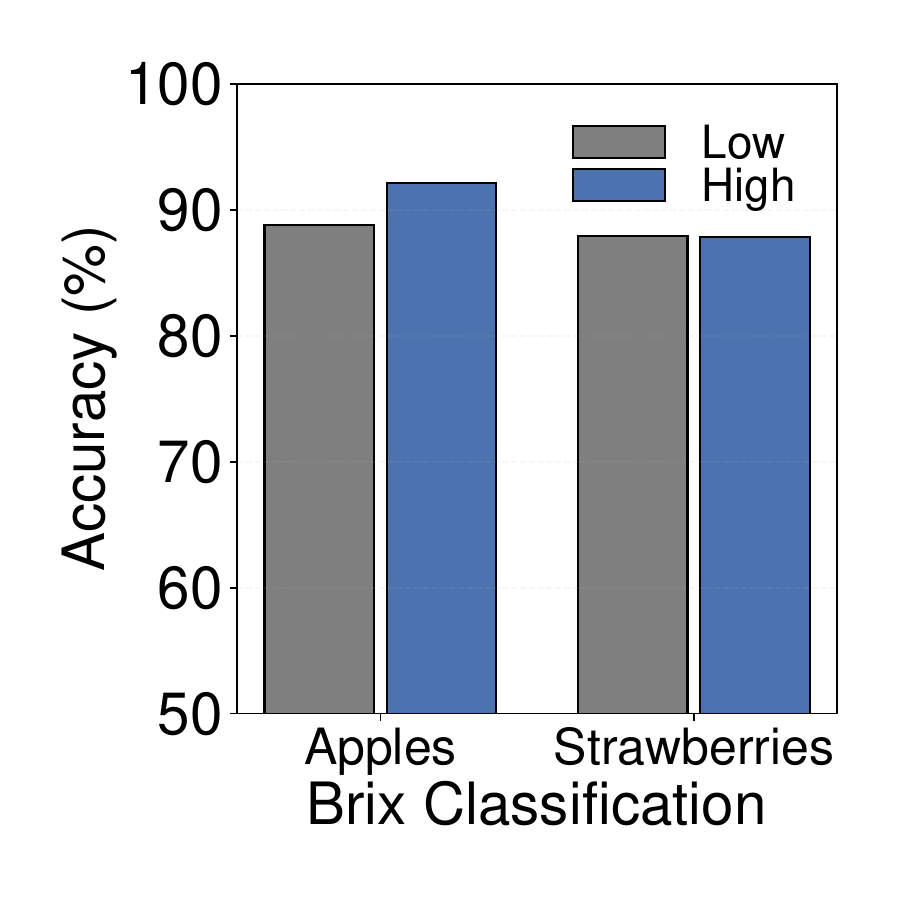}}
  \hspace*{0.2em}
  \subfloat[SF-Net\label{fig:resultsbredictionperformance}]{\includegraphics[trim={0cm 0.2cm 0.8cm 0cm},clip,height=0.3\linewidth]{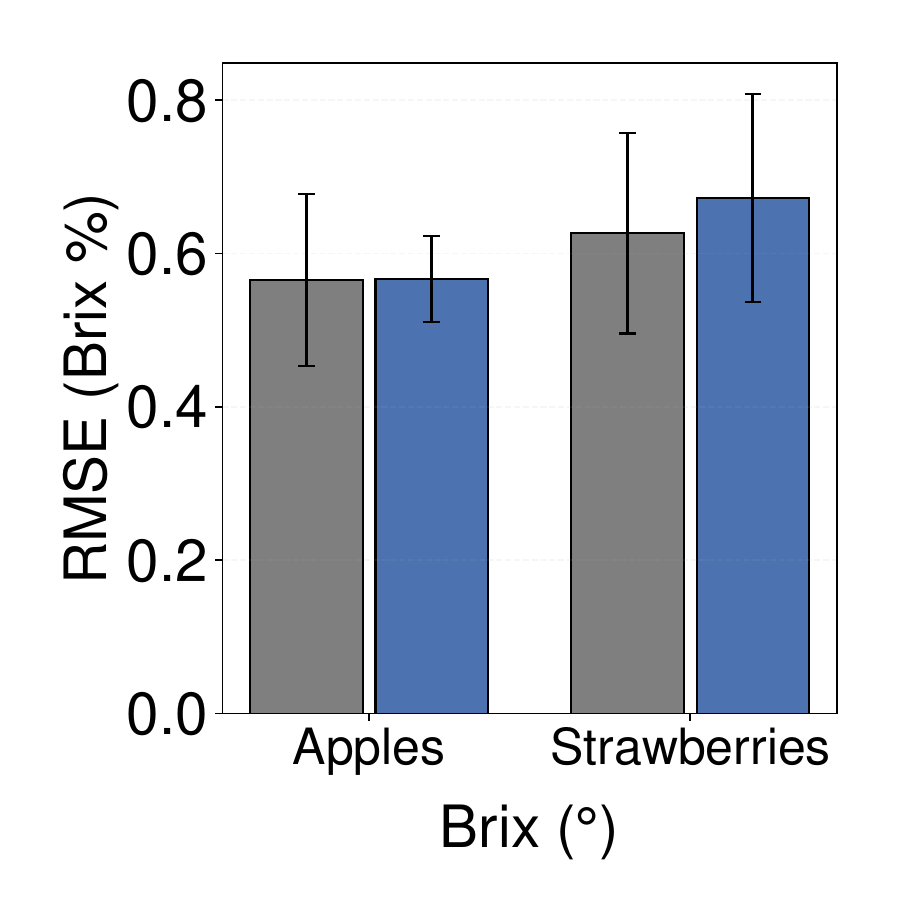}}
  \caption{Experimental setup and system performance of SF-PointNet (Stage~1) and SF-Net (Stage~2).}
  \label{fig:resultsoutputmodel}
  \ifdefined\USEACMSTYLE
    \Description{}
  \fi
\end{figure}

\subsubsection{Impact of Spatial Radius Window Size in ToF-Based Classification}

\begin{figure}[htbp]
  \centering
  \subfloat[Apple ToF segmentation with various radius\label{fig:resultspointcloudradiussegment}] {\includegraphics[trim={0cm 2.5cm 0.2cm 0.0cm},clip,height=0.45\linewidth]{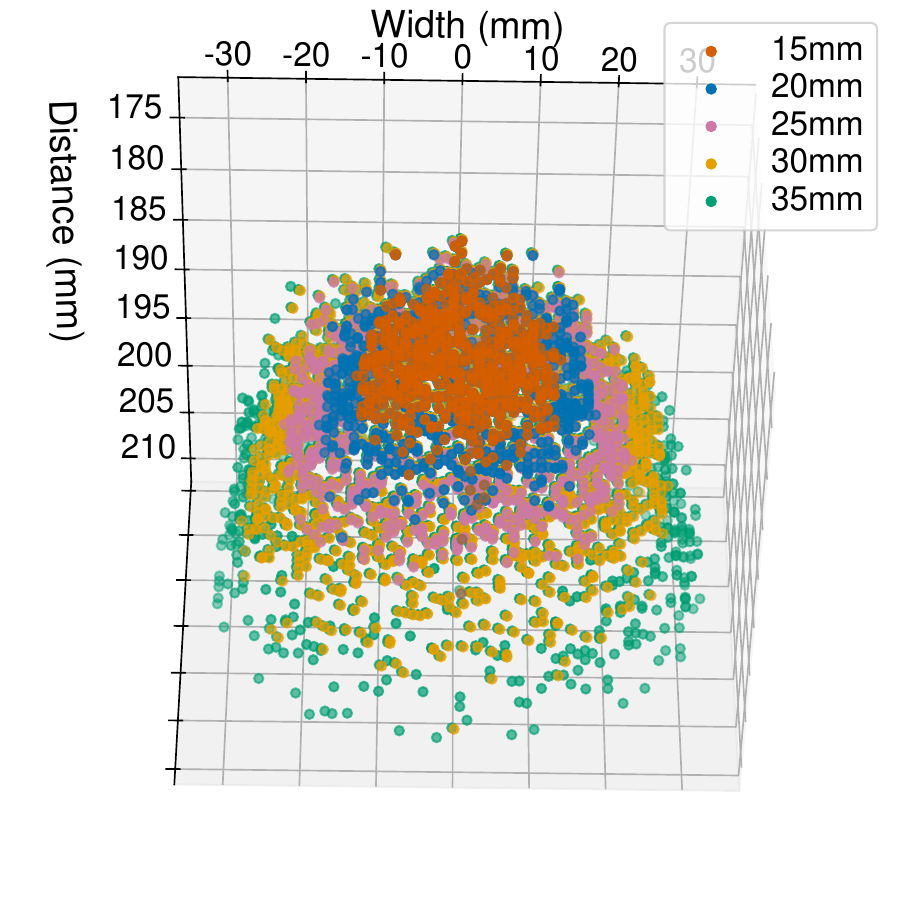}}
  \hspace*{0.2em}
  \subfloat[Accuracy vs Apple Radius\label{fig:resultspointradiusclassificationaccuracy}]{\includegraphics[height=0.41\linewidth]{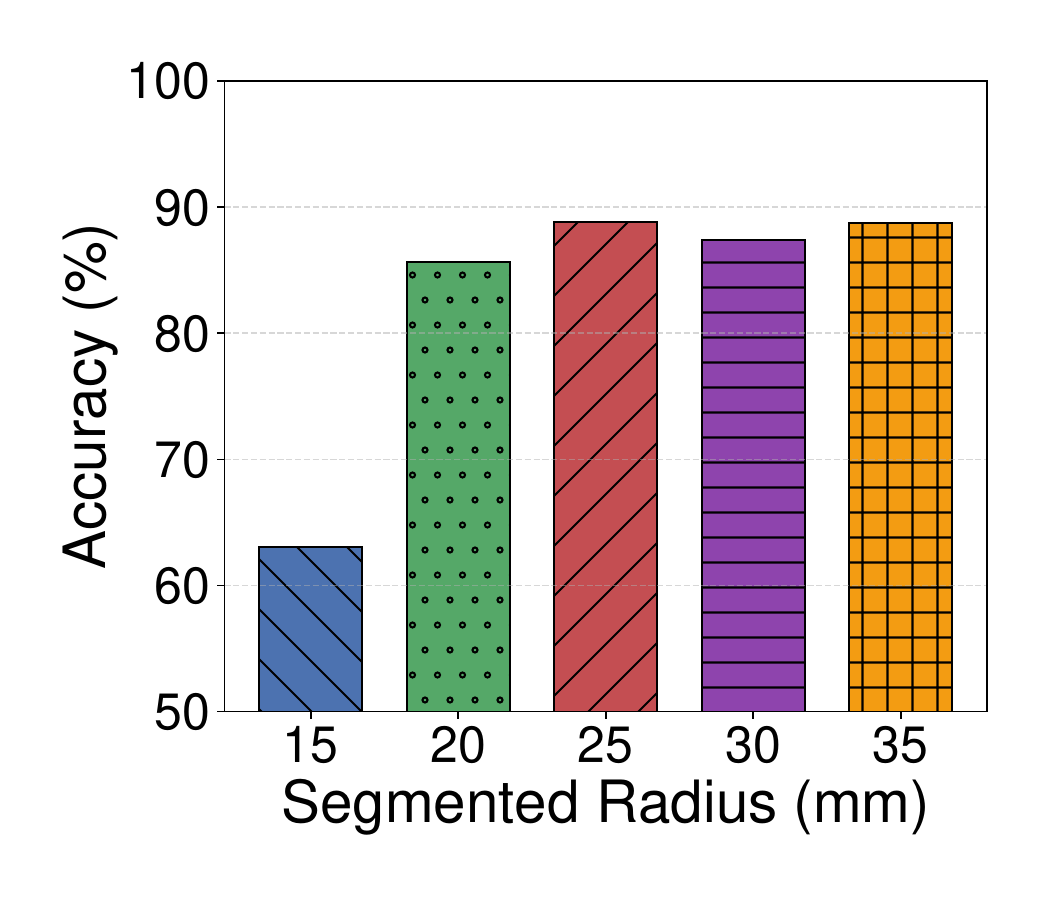}}
  \caption{Stage 1 classification accuracy versus spatial radius window size for point cloud sampling. The optimal radius (~25 mm) balances accuracy and detail preservation.}
  \ifdefined\USEACMSTYLE
    \Description{Classification accuracy at different spatial radius sizes in Stage 1.}
  \fi
\end{figure}

To evaluate the classification performance at Stage 1, extensive testing analyzed the influence of spatial radius parameters used in point cloud sampling on classification accuracy. Using ToF data collected from the apple data, performance metrics were assessed across radius windows ranging from 15 mm to 35 mm, summarized in Fig.~\ref{fig:resultspointradiusclassificationaccuracy}. Accuracy improved up to an optimal radius of approximately 25 mm, with a classification accuracy of 89\%, effectively guiding Stage 2 activation and significantly reducing unnecessary close-range scans.

\begin{table}[ht!]
\centering
\caption{SSC (\textdegree Brix) properties of fruit samples.}
\begin{tabular}{|c|c|c|c|c|}
\hline
\textbf{Items} & \textbf{Max} & \textbf{Min} & \textbf{Mean} & \textbf{SD} \\ \hline
Apple & 14.6 & 9.9 & 12.3 & 1.2 \\ \hline
Strawberry & 10.16 & 6.26 & 7.9 & 1.10 \\ \hline
\end{tabular}
\label{tab:sugar_content_apple_samples}
\end{table}

\subsection{Stage 2 SF-Net Regression Performance}

The predictive performance of \textit{SF-Net} for estimating fruit sugar content is shown in Figure~\ref{fig:resultsbredictionperformance}, with baseline comparisons summarized in Table~\ref{tab:mae_results}. SF-Net achieves a RMSE of 0.57~\textdegree Brix on apples, outperforming the strongest baseline (multivariate linear regression at 0.73~\textdegree Brix) by approximately 22\%. This improvement demonstrates the benefit of incorporating depth-encoded features into the regression pipeline and validates the end-to-end design of the two-stage system.

\revisedtext{
As illustrated in Figure~\ref{fig:resultsbredictionperformance}, the RMSE for strawberries is slightly higher than for apples, with larger variance across samples. This result is expected, given that the strawberry regression model was trained using only NIR spectral data, without depth input. The relatively tighter error bounds for apples confirm the stabilizing effect of depth integration, while the wider variance for strawberries highlights the need for better pose control or additional multimodal sensing.}

\revisedtext{Due to hardware constraints, full multimodal integration was conducted only for apple samples. The strawberry experiments relied solely on the NIR spectrometer at a working distance of approximately 3 cm, which falls below the minimum range of the ToF sensor. As a result, strawberry sugar estimation was performed using a reduced version of SF-Net without depth input, serving as a preliminary baseline. Extending full two-stage sensing to strawberries is planned for future work and will require short-range depth sensors compatible with high-resolution fruit surfaces.}

\begin{table}[ht!]
\centering
\caption{Comparison of SF-Net with SOTA Methods}
\begin{tabular}{|c|c|c|c|}
\hline
Related Works                           & Method    & Need Contact      & RMSE (\textdegree) \\ \hline
Zhao et al.~\cite{zhao_flight_2023}     & MLR       & Yes               & 0.73  \\ \hline
Tran et al.~\cite{tran_portable_2020}   & PLSR/MLR  & Yes               & 0.73  \\ \hline
\sn                                     & SF-NET    & No                & 0.57  \\ \hline
\end{tabular}
\label{tab:mae_results}
\end{table}

\subsubsection{Ablation Study}

\begin{figure}[htbp]
  \centering
  \subfloat[Ablation Test\label{fig:resultssfnetablationmoduleblocksrmse}]{\includegraphics[height=0.32\linewidth]{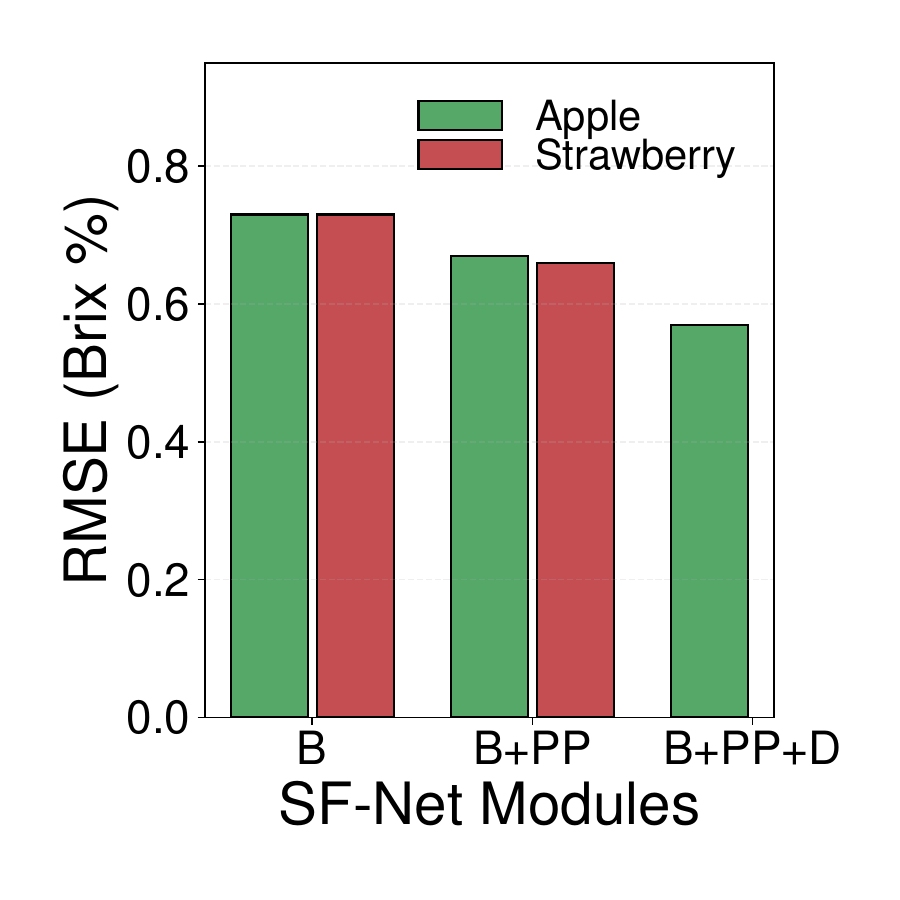}}
  \hspace*{0.1em}
  \subfloat[Raw NIR Intensity\label{fig:resultsappledistanceevaluation}]{\includegraphics[height=0.30\linewidth]{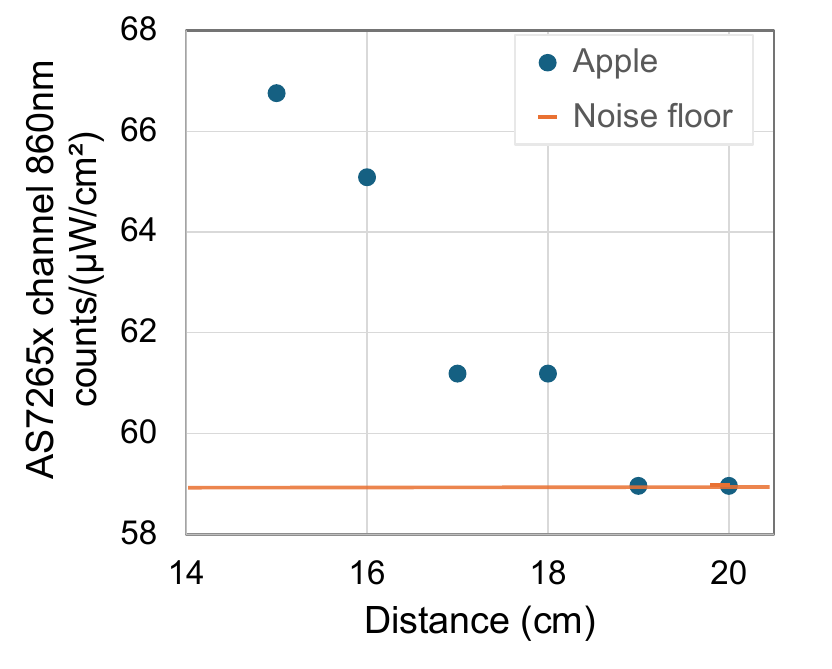}}
  \hspace{0.2em}
  \subfloat[Distance Test\label{fig:resultsdistancetestrmse}]{\includegraphics[height=0.32\linewidth]{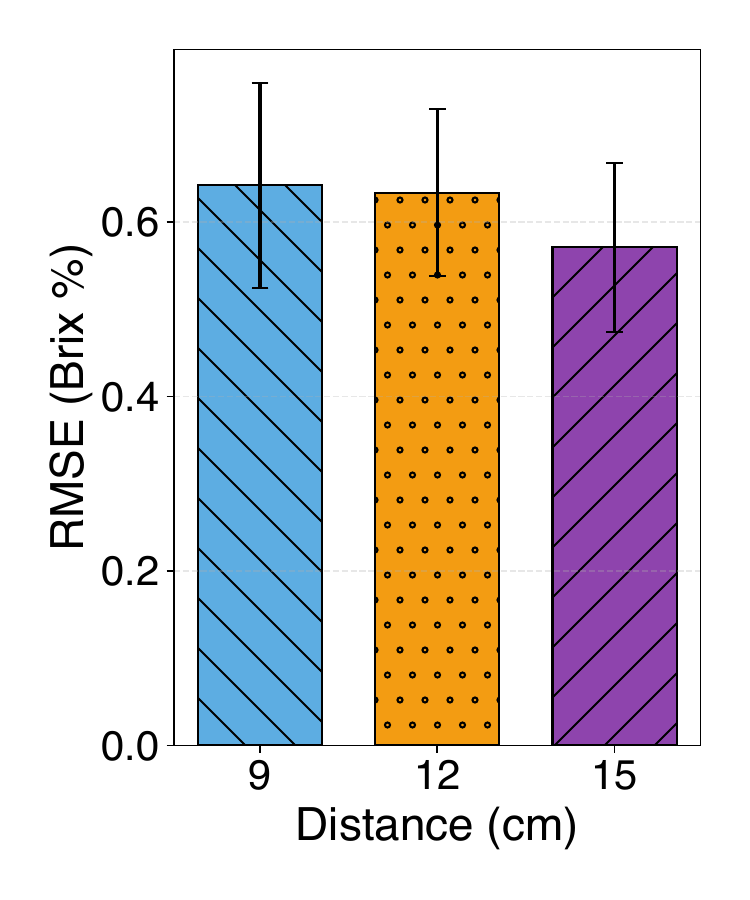}}
  \caption{Evaluation of SF-Net components (Base 'B' Preprocessing 'PP' Depth 'D') and impact of fruit distance on prediction accuracy.}
  \label{fig:resultsablationsstudyandappledistance}
  \ifdefined\USEACMSTYLE
    \Description{}
  \fi
\end{figure}

We conducted an ablation study to isolate the impact of key components within the SF-Net regression pipeline, including spectral preprocessing, depth encoding, and multimodal sensor fusion. The results are shown in Figure~\ref{fig:resultssfnetablationmoduleblocksrmse}.

\paragraph{Impact of Spectral Preprocessing}
The application of SNV normalization was critical by removing multiplicative and additive effects due to scattering and baseline shifts, ensuring that variations in the spectral data were primarily attributable to compositional differences rather than external factors. Removing the SNV normalization step led to an increase in RMSE of approximately 8\%, highlighting the necessity of this preprocessing stage for reliable sugar content prediction.

\paragraph{Incorporation of Depth Vector}
The inclusion of a depth vector as an additional feature improved prediction accuracy by enabling the model to account for potential surface variability (e.g., skin thickness and texture). Adding the depth vector reduced the RMSE by roughly 15\%, from 0.67° Brix to 0.57° Brix, which strengthens the argument for integrating depth information to mitigate artifacts arising from surface irregularities.

\paragraph{Sensor Fusion}
The combined use of spectral and depth features outperformed spectral-only models, indicating that the two modalities offer complementary information. Spectral signals capture subsurface chemical absorption, while depth data improves robustness by contextualizing surface structure and helping correct for external confounding factors like lighting and distance.

These results affirm that each module in the SF-Net architecture contributes meaningfully to system accuracy and robustness. Future work may explore alternative depth encoding techniques or lightweight attention mechanisms for adaptive fusion to further enhance performance.

\subsubsection{Impact of Apple-Sensor Distance}

Figure~\ref{fig:resultsappledistanceevaluation} illustrates the raw intensity data captured at 860 nm from the AS7265x spectral sensor from an apple at varying distances, ranging from 15 cm to 20 cm. At distances beyond 18 cm, signal strength approaches the noise floor of the AS7265x spectrometer, leading to degraded spectral quality. Based on these findings, we selected a target measurement distance of 15 cm with a ±2 cm tolerance, balancing signal integrity and practical usability in field conditions.

Figure~\ref{fig:resultsdistancetestrmse} shows the performance of SF-Net across three different distances. The lowest RMSE error occurred at 15 cm, reflecting the system's training focus on this range. At distances of 12 cm and 9 cm, the RMSE increased, suggesting a decline in prediction accuracy as the system moved away from its optimal measurement range. This limitation highlights the need for additional training data across a broader range of distances to improve the model’s generalization and performance in diverse scenarios.

\subsubsection{Impact of Multiple Fruit Objects}
\begin{figure}[htbp]
  \centering
  \subfloat[Crowded apples\label{fig:resultstwoapplesviewingangle}]{\includegraphics[height=0.42\linewidth]{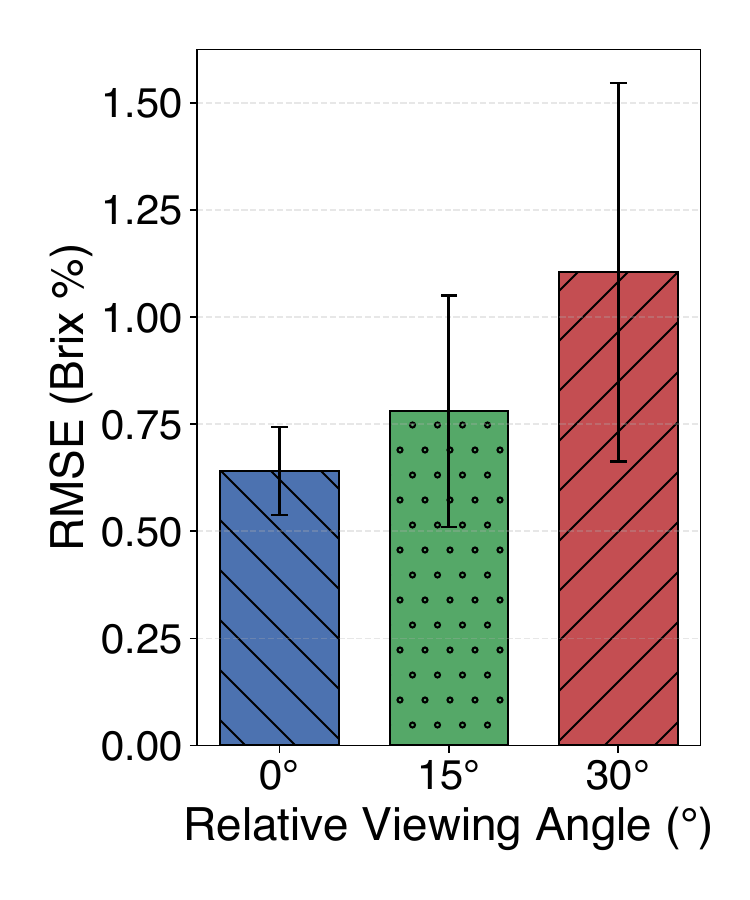}}
  \hspace*{0.5em}
  \subfloat[Indoor/moonlight light $\Delta \%$ 
  \label{fig:resultsapplelightingevaluation}]{\includegraphics[trim={0.3cm 0cm 0.4cm 0.2cm},clip,height=0.37\linewidth]{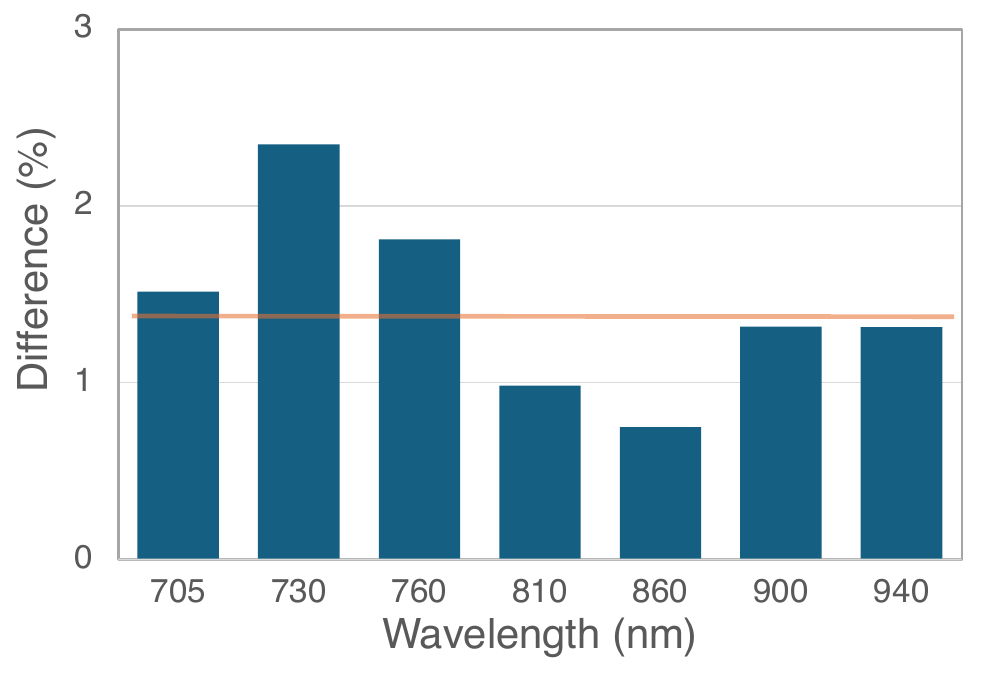}}
  \caption{Evaluation of multiple apples and experiment lighting conditions}
  \label{fig:resultsappledistanceandlightingevaluation}
  \ifdefined\USEACMSTYLE
    \Description{}
  \fi
\end{figure}

\revisedtext{To simulate realistic deployment in orchards or packing lines, we evaluated SweetFruit’s performance in the presence of multiple apples in the field of view. In this setup, one apple served as the primary target and was positioned directly in front of the sensor, while a second apple was placed at various angles relative to the first. As shown in Figure~\ref{fig:resultstwoapplesviewingangle}, increasing the viewing angle of a secondary apple leads to partial occlusion and spectral contamination, which in turn increases RMSE.}

At a relative viewing angle of 0º, the primary apple occupied the majority of the sensor's field of view, resulting in minimal interference and the lowest RMSE error. As the relative viewing angle increased, the secondary apple became more visible, introducing spectral noise and disrupting the system's ability to isolate the primary apple's features. This interference progressively increased the RMSE error, demonstrating the system's sensitivity to overlapping objects. To mitigate this error, the sensor should be positioned to focus on the nearest object and reduce visually overlapping fruit surfaces. This would ensure reliable measurements in real-world scenarios involving multiple fruits.

\subsubsection{Impact of Ambient Lighting Conditions}

Given the potential variability in lighting conditions, evaluating the impact of different environments on spectral measurements was important. Two real-world deployment scenarios were considered, one designed to operate on an orchard at night using an automated robotic setup, and one designed to work indoors on a conveyor belt using controlled lighting. The spectral differences in the NIR band between indoor artificial lighting (used during experiments) and moonlight in an open field were analyzed.
Figure~\ref{fig:resultsapplelightingevaluation} shows the normalized differences between the indoor-light and moonlight spectra, expressed as a percentage of the target apple spectrum. The percentage difference is calculated as follows:

\[
\text{Difference \%} = \frac{\text{Indoor Light} - \text{Moonlight}}{\text{Target Spectrum}} \times 100
\]

Preliminary comparisons between indoor artificial lighting and moonlight conditions indicate a spectral offset; however, the effect on prediction accuracy appears minimal and within the operating range of the AS7265x sensor, particularly around the key 860 nm absorption band. Future work will involve a more systematic study under various ambient lighting conditions and the potential integration of adaptive lighting compensation techniques.

\subsection{Discussion}
\revisedtext{
The \snn system demonstrated that practical fruit quality sensing can be achieved using a structured sensing pipeline and low-cost components. By decoupling the classification and regression stages, the system reduced the inference time and resource usage, while maintaining high accuracy. This modular architecture would suit deployment on embedded or mobile platforms, where compute and energy budgets are limited.

One key advantage of the two-stage design is that it avoided the need to process close-range time-consuming spectral data measurements for every fruit. This not only improved the throughput but also allowed selective sensor activation based on task requirements, which can be especially useful in energy-constrained settings such as field robots or handheld devices. While this version used fixed thresholds for stage transition, future systems could incorporate dynamic policies driven by uncertainty estimation or deployment context.

Another important takeaway is the benefit of task-specific sensor selection over full multimodal fusion. Rather than integrating all available sensor data in every step, \snn shows that low-cost sensors can be used efficiently when each is aligned with a specific subtask-geometry for classification and spectral response for sugar estimation.

Although \snn was evaluated on only two fruit types (green apples and strawberries), its modular sensing architecture is designed with generalizability in mind. Prior work such as VNiScan-Fruit~\cite{wang_vniscan-fruit_2026} has shown that NIR-based Brix estimation can effectively transfer across a wide range of fruits with appropriate calibration. Our results reinforce this potential: the classification stage operates on geometric cues that generalize across fruit morphologies, while the spectral regression stage can be retrained for new fruit types with minimal hardware changes. This supports future extensions to broader fruit categories and mixed-sample environments.

Finally, \snn bridges a gap between mobile systems research and real-world agricultural sensing needs. Many existing solutions in this space either prioritize high-end accuracy with expensive hardware or sacrifice robustness for simplicity. Our results suggest that structured, resource-aware pipelines offer a promising alternative for scalable, field-deployable sensing applications in agriculture and beyond.}

\subsection{Limitations}
\revisedtext{
While \snn demonstrates promising results for low-cost, non-destructive fruit sugar estimation, several limitations remain.

First, the system's performance on strawberries is constrained by the current sensing configuration. Due to their irregular shape and small size, the ToF sensor often struggles to capture clean surface profiles, and NIR measurements are more sensitive to alignment and surface variability. Our current results for strawberries rely only on close-range NIR sensing, without full two-stage integration.

Second, the system requires fruit to be relatively still during capture, limiting deployment in fast-moving environments such as conveyor belts or robotic arms without motion compensation. While our current prototype supports real-time inference, fully integrating the system into dynamic workflows would require additional hardware or motion tracking mechanisms.

Third, our evaluation was limited to two fruit types with controlled lab lighting. Broader validation under diverse environmental conditions, such as varying ambient light, humidity, or fruit surface states (e.g., wet, dusty), is necessary to ensure generalizability in the field.

Lastly, while the two-stage pipeline is effective for reducing sensing and computation, it assumes a fixed decision boundary between coarse classification and fine regression. Adaptive or confidence-aware transitions between stages could further improve efficiency and accuracy, especially when deployed on resource-constrained platforms.

Addressing these limitations will be essential for scaling SweetFruit to a wider range of real-world agricultural and consumer-facing scenarios.}

\section{Related Work}\label{sec:relatedwork}
\ifdefined\USEIEEESTYLE \noindent \fi
\snn is related to a wide-range of research areas and we group the related works into two key areas consisting of: (i) \nameref*{itm:rw_1} and (ii) \nameref*{itm:rw_2}.

\subsection{Fruit sugar content estimation}\label{itm:rw_1}

Accurately measuring sugar content in fruits is essential for assessing ripeness, shelf life, and market value. Traditional approaches such as HPLC offer precise results but require destructive sampling and laboratory infrastructure, making them unsuitable for high-throughput or in-field scenarios~\cite{smithson_investigating_2021}.

Recent work has explored non-contact sensing techniques using electromagnetic waves, including millimeter-wave and radio-frequency (RF) sensing. For example, Yang et al. demonstrated sugar estimation using 60 GHz millimeter-wave signals, achieving correlation coefficients up to 85\%~\cite{yang_feasibility_2019}. Tavasoli et al. introduced \textit{SugarWave}, which also uses millimeter-wave technology to estimate sugar content, reporting a median error of 1.4° Brix~\cite{tavasoli_sugarwave_2023}. These approaches, while promising, often struggle with scalability or require specialized equipment, highlighting the need for cost-effective, flexible solutions.

\revisedtext{Smartphone-based approaches such as MobiChem~\cite{li_mobichem_2025} leverage commodity front-facing cameras and screen illumination to perform multispectral imaging. These systems demonstrate fine-grained sugar classification on consumer hardware but depend on narrowband optical filters or per-device calibration to maintain consistency. In contrast, our design uses a compact NIR spectrometer plus a ToF camera to enable deeper subsurface sensing at NIR wavelengths (700--1000\,nm), which are more strongly correlated with sugar absorption bands (880--940\,nm).}

\subsection{NIR spectroscopy for non-destructive sensing}\label{itm:rw_2}

NIR spectroscopy has emerged as a powerful tool for non-destructive quality assessment in fruits, capable of evaluating parameters such as sugar content, firmness, and acidity. Early work by Kawano et al. validated the use of NIR spectroscopy for Brix prediction, achieving high correlation scores (R² = 0.97) in apples~\cite{kawano_determination_1992}.

Recent work has focused on portable and embedded NIR solutions. The AS7265x sensor, used in this work, supports 18 discrete VIS+NIR bands in a compact package and has been deployed in mobile and UAV-based sensing platforms~\cite{ams_as7265x_2024}. Tran et al. demonstrated that low-cost spectrometers can match lab-grade performance with RMSE as low as 0.4° Brix \cite{tran_portable_2020}, and Zhao et al. deployed a UAV-mounted NIR sensor achieving 0.52° Brix RMSE \cite{zhao_flight_2023}. 

Advanced preprocessing techniques have also been introduced to improve the robustness of NIR systems. Methods such as SNV correction and competitive adaptive reweighted sampling (CARS) enhance model accuracy by reducing noise and selecting optimal wavelengths \cite{huang_stacking_2024, zeng_research_2024}. Vaudelle et al. examined how physical properties like fruit size and skin thickness affect NIR penetration, informing calibration strategies for reliable field use \cite{vaudelle_influence_2015}.

Finally, multimodal approaches that combine structural and spectral data have shown promise in mitigating surface irregularities and spatial variability. For example, Medic et al. employed multispectral LiDAR to fuse 3D geometry with spectral reflectance, enabling more accurate prediction of internal sugar and dry matter content~\cite{medic_remotely_2024}. Similarly, Wang et al. proposed \textit{VNiScan-Fruit}, a smartphone-based VIS-NIR system that generalizes Brix estimation across six fruit types using a unified regression model~\cite{wang_vniscan-fruit_2026}. This work highlights the feasibility of multi-fruit generalization with proper spectral coverage and supports SweetFruit’s potential beyond its initial fruit set.

\section{Conclusion and Future Work}\label{sec:conclusion}
\ifdefined\USEIEEESTYLE \noindent \fi

We presented \sn, a mobile, low-cost sensing system for non-destructive estimation of fruit sugar content using a structured two-stage pipeline. By combining a lightweight depth-based classifier with a compact NIR-based regression model, \snn enables high-throughput, real-time prediction of Brix values using only off-the-shelf hardware. Our evaluation on apples and strawberries demonstrates that this architecture offers strong generalization, high screening accuracy (over 90\%), and precise sugar estimation with a root mean square error of 0.57~\textdegree Brix, improving upon single-sensor baselines while keeping sensing and computational requirements low. Our design complements recent work such as VNiScan-Fruit~\cite{wang_vniscan-fruit_2026}, which demonstrates generalizable sugar estimation across diverse fruits.

Unlike systems that rely on complex sensor fusion or expensive modalities, \snn demonstrates that practical and modular sensing strategies can be effectively tailored to agricultural quality assessment. The system’s two-stage design also enables flexible deployment on resource-constrained platforms such as handheld devices, drones, or autonomous pickers.

For future work, we plan to extend \snn to a wider range of fruit varieties and investigate its performance in field environments under varying lighting and temperature conditions. We also aim to explore adaptive sensor activation and further optimize runtime performance for edge deployment. These results suggest that structured, low-cost sensing pipelines can serve as a foundation for scalable, real-time food quality analysis and broader mobile sensing applications.

\vspace{\baselineskip}

\begin{acks}
    This work is supported in part by the Future Food Systems under Grant Fund RE932. 
    This research includes computations using the computational cluster Katana supported by Research Technology Services at UNSW Sydney.

    Statement: During the preparation of this work the author used OpenAI's GPT-5 to improve the readability of various sections. After using this tool/service, the author reviewed and edited the content as needed and take full responsibility for the content of the publication.

\end{acks}

\bibliographystyle{ACM-Reference-Format}
\bibliography{bibliography/ch6-sugar}

\end{document}